\documentclass[aps,twocolumn,showpacs,amsmath,amssymb,superscriptaddress,prc]{revtex4-2}
\usepackage{graphicx,here}
\usepackage{multirow}
\usepackage{tikz}
\usepackage{epstopdf}
\usepackage[percent]{overpic}
\usepackage{scrextend}
\usepackage{xcolor,xspace}
\usepackage[ulem=normalem]{changes}
\usepackage{hyperref}
\hypersetup{colorlinks=True,urlcolor=blue,linkcolor=blue,citecolor=blue,filecolor=black}

\colorlet{Changes@Color}{red}
\usepackage{amsmath}
\usepackage{dcolumn,color,footnote,bm,braket}
\usepackage{url,longtable,tabularx}
\usepackage{threeparttable}

\usepackage{fancybox}
\usetikzlibrary{matrix}

\newcommand\+{\dagger}
\newcommand\jr{j_\rho}

\newcommand\tm{\hat T^{(M1)}}
\newcommand\tmb{\hat T_{\mathrm{B}}^{(M1)}}
\newcommand\tmf{\hat T_{\mathrm{F}}^{(M1)}}
\newcommand\te{\hat T^{(E2)}}
\newcommand\teb{\hat T_{\mathrm{B}}^{(E2)}}
\newcommand\tef{\hat T_{\mathrm{F}}^{(E2)}}

\begin{document}

\title{Questioning the wobbling interpretation of low-spin bands in $\gamma$-soft nuclei within the interacting boson-fermion model}

\author{K. Nomura}
\email{knomura@phy.hr}
\affiliation{Department of Physics, Faculty of Science, 
University of Zagreb, HR-10000, Croatia}

\author{C. M. Petrache}
\affiliation{Universit\'{e} Paris-Saclay, CNRS/IN2P3, IJCLab, 91405 Orsay, France}

\date{\today}

\begin{abstract}
An alternative interpretation of 
the recently reported low-lying excited bands 
in $\gamma$-soft odd-mass nuclei as  
wobbling bands is presented in terms of the 
interacting boson-fermion model. 
The model Hamiltonian is determined 
based on the mean-field 
calculations with the nuclear energy density 
functionals. The predicted mixing 
ratios of the $\Delta{I}=1$ electric quadrupole 
to magnetic dipole  transition 
rates between yrast bands and those yrare bands 
previously interpreted as wobbling bands 
in $^{135}$Pr, $^{133}$La, $^{127}$Xe, 
and $^{105}$Pd 
are consistently smaller in magnitude than 
the experimental values on which 
the wobbling interpretation is based. 
These calculated mixing ratios indicate 
predominant magnetic character in agreement 
with the new experimental data. 
The earlier 
wobbling assignments are severely questioned. 
\end{abstract}

\maketitle

\section{Introduction}

Ground-state shape of most nonspherical 
nuclear systems is characterized by 
axially-symmetric quadrupole deformation 
\cite{BM_II}. 
The axial symmetry, 
i.e., invariance under rotation about the 
symmetry axis of the intrinsic frame, 
is, however, broken in many nuclei. 
The nonaxial nuclear shapes as well as 
the resulting triaxially deformed rotors 
are a prominent feature of nuclear structure. 
A fingerprint of the rigid triaxiality 
is wobbling motion \cite{BM_II}, 
a collective mode 
in which the principal axis of 
a triaxial rotor corresponding to the 
largest moment of inertia oscillates 
about the space-fixed angular momentum. 
The phenomenon has attracted much attention 
in nuclear physics, and is also recognized 
in finite many-body microscopic and macroscopic systems in general.

The wobbling motion in nuclei can be identified 
experimentally through 
the observation of rotational 
bands that are connected to each other by 
predominant $\Delta{I}=1$ electric quadrupole 
($E2$) transitions, 
because the collective oscillation 
of the entire nuclear charge is involved. 
Traditionally, excited bands that manifest 
features of wobbling motion have been 
identified in high-spin bands of the 
odd-mass Lu and Ta nuclei in the mass $A\approx160$ region 
\cite{wobblingprl-2001,wobblingprl-2002,165Lu,167Lu,161Lu,167Ta}. 
More recent experiments have shown new evidence 
for wobbling bands in odd-mass nuclei 
in several other mass regions, observed  
in the low-spin regime, 
e.g., in 
$^{135}$Pr \cite{matta2015,sensharma2019}, 
$^{133}$La \cite{biswas2019}, 
$^{105}$Pd \cite{timar2019}, 
$^{127}$Xe \cite{CHAKRABORTY2020}, 
$^{187}$Au \cite{sensharma2020}, and
$^{183}$Au \cite{nandi2020},
as well as 
at medium spins in $^{130}$Ba \cite{chen-2019} 
and $^{136}$Nd \cite{chen-2021}.
In comparison to the high-spin wobbling bands 
of strongly-deformed triaxial nuclei in 
the mass $A\approx 160$ region, the new experiments 
have proposed the occurrence of low-spin wobbling 
motion in normal-deformed $\gamma$-soft nuclei, 
which are characterized by a collective potential 
that is soft in nonaxial deformation and has small 
quadrupole deformation. 
The search for new regions of wobbling motion 
expands the frontier of nuclear 
collective motion, but should be accompanied 
by increasing experimental rigour.

In fact, it is of crucial importance 
to critically assess the reported experimental evidence 
for the wobbling bands. 
The wobbling interpretation requires 
connecting transitions with predominant 
electric character, which can be established 
by extracting mixing ratios $\delta(E2/M1)$
 with magnitudes larger than 1, which signifies predominance of the electric over the magnetic components.
Actually, new experiments that involve 
angular distribution combined with 
linear polarization measurements 
on excited bands in $^{187}$Au \cite{Guo} 
and $^{135}$Pr \cite{135Pr-Lv} 
showed that the interband 
transition between the proposed wobbling bands 
and the yrast bands in these nuclei are predominantly 
magnetic. 
The wobbling interpretation of 
the newly found bands has been mostly 
based on 
a particle-rotor picture, in which the 
configuration that embodies the wobbling motion 
is explicitly considered within the intrinsic 
frame of reference   
\cite{Transverse-2014,matta2015,sensharma2019,biswas2019,timar2019,135Pr-Lv}. 
On the other hand, it would be useful to 
give an alternative 
theoretical interpretation of the proposed 
low-spin wobbling bands if the character of the connecting transitions is not predominantly electric, as required by the collective wobbling motion.

In this paper, we shall 
consider the recently proposed low-spin wobbling 
bands of $\gamma$-soft odd-mass nuclei 
$^{135}$Pr, $^{133}$La, $^{127}$Xe, and $^{105}$Pd, 
within the interacting boson-fermion model 
(IBFM) \cite{iachello1979,IBFM}, 
with the Hamiltonian 
determined by the constrained mean-field 
calculations that are based upon the nuclear energy 
density functional (EDF) 
\cite{bender2003,vretenar2005,robledo2019,schunk2019}. 
The aim of this work is to provide an alternative 
interpretation of the observed non-yrast bands of 
the above odd-mass nuclei. 
Our calculation reproduces the new data on $^{135}$Pr 
\cite{135Pr-Lv} and the old 
data on $^{105}$Pd \cite{rickey1977}, 
but is in contradiction 
with those experimental data on which 
the wobbling interpretation is based.
Here we mainly focus on the excitation spectra 
and electromagnetic transition properties 
of these non-yrast bands, 
that are obtained from the diagonalization of the 
IBFM Hamiltonian in the laboratory frame 
of reference. 
We also note that certain intrinsic properties of 
odd-mass nuclei can be dealt with within the IBFM 
framework as well, by making use of the formalism of 
coherent state that is generalized to coupled 
boson-fermion systems 
\cite{leviatan1988,alonso1992,IBFM}. 
This procedure has been extensively used for analyzing 
properties in the intrinsic frame, 
including the studies of the quantum shape-phase 
transitions in odd-mass nuclei 
\cite{iachello2011,boyukata2021}.

The paper is organized as follows. 
In Sec.~\ref{sec:theory}, we outline 
the theoretical procedure to construct the 
IBFM Hamiltonian based on the mean-field 
calculations. 
Section~\ref{sec:results} shows our results 
including the excitation spectra, $E2/M1$ mixing ratios, 
$B(E2)$ and $B(M1)$ transitions for the 
considered odd-mass nuclei $^{135}$Pr, 
$^{133}$La, $^{127}$Xe, and $^{105}$Pd. 
Summary of the main results is given 
in Sec.~\ref{sec:summary}.

\section{Theoretical procedure\label{sec:theory}}

%
%
\begin{figure}[ht]
\begin{center}
\includegraphics[width=\linewidth]{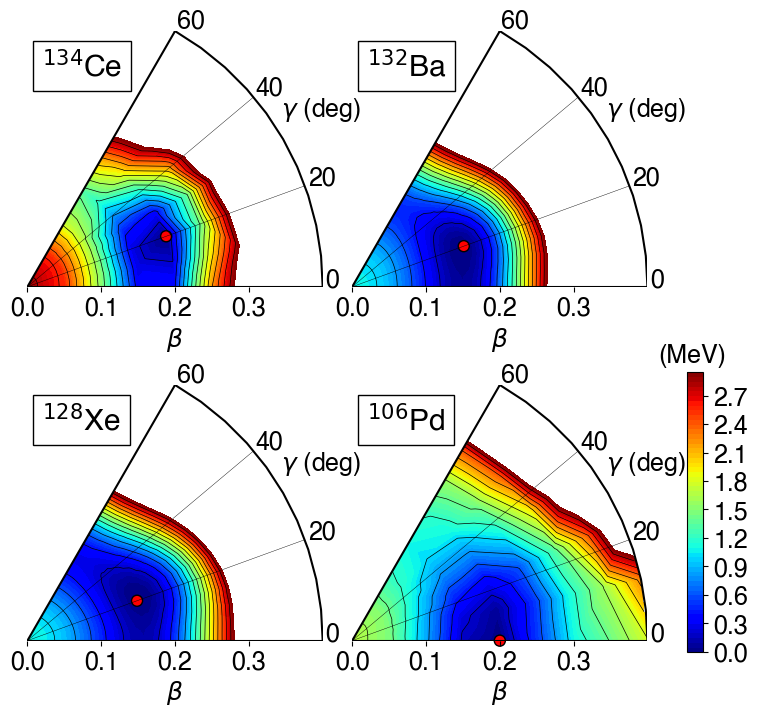}
\caption{Potential energy surfaces (PESs) 
for the even-even core nuclei obtained 
from the mean-field calculations with quadrupole 
degrees of freedom $\beta$ and $\gamma$. 
Two representative nuclear effective 
interactions are employed: Gogny D1M \cite{D1M} 
(for $^{132}$Ba, and $^{128}$Xe) 
and DD-PC1 
\cite{DDPC1} (for $^{134}$Ce and $^{106}$Pd) 
EDFs. 
The total mean-field energies 
are plotted up to 3 MeV, normalized with 
respect to the global minimum which is represented 
by a solid circle. 
The energy difference between the 
neighboring contours is 0.2 MeV. Note the PESs 
for $^{132}$Ba and $^{128}$Xe are taken from 
Ref.~\cite{nomura2017odd-3}. 
}
\label{fig:pes}
\end{center}
\end{figure}

In even-even nuclei, to a good approximation,  
nucleons are coupled pairwise 
and the presence of such pairs 
play an important 
role in nuclear dynamics, determining basic 
parameters of vibrational and rotational spectra. 
In odd-mass nuclear systems, 
one has to consider explicitly the
unpaired nucleon and 
treat the collective and noncollective 
(single-particle) degrees of freedom 
on the same footing \cite{bohr1953}. 
A major assumption in the present 
work is that the low-lying states of 
an even-even nucleus is described by the 
interacting boson model (IBM) \cite{IBM}, 
consisting of the monopole 
$s$ (with spin and parity $L=0^{+}$) 
and quadrupole $d$ ($L=2^{+}$) bosons, which 
represent the collective $S$ and $D$ pairs 
of valence nucleons, respectively. 

The low-energy structure of a given odd-mass 
nucleus is determined by the interaction between 
an odd fermion and the even-even boson (IBM) core. 
Specifically,  
$^{135}$Pr ($^{133}$La) is a system composed 
of even-even 
core $^{134}$Ce ($^{132}$Ba) plus an odd proton 
particle, while $^{127}$Xe ($^{105}$Pd) 
is composed of 
the $^{128}$Xe ($^{106}$Pd) even-even core coupled 
to the odd neutron hole. 
For all these odd-mass 
nuclei, the fermion space 
corresponds to the proton $Z$ or neutron 
$N=50-82$ major oscillator shell, 
hence only the orbital $1h_{11/2}$ 
is considered to describe negative-parity states. 
In general, the IBFM Hamiltonian is given by
\begin{align}
\label{eq:ham}
 \hat H 
= \hat{H}_\text{B} 
+ \hat{H}_\text{F}
+ \hat{V}_\text{BF},
\end{align}
where $\hat{H}_\mathrm{B}$ stands for 
the IBM Hamiltonian for an even-even core, 
$\hat{H}_\mathrm{F}$ is the 
single-nucleon Hamiltonian, and  
$\hat{V}_\mathrm{BF}$ represents 
the boson-fermion interaction. 

In the first step of the present theoretical 
analysis, we carry out, for each even-even core nucleus, 
the constrained mean-field calculations 
\cite{RS} based on a given EDF, 
and obtain the potential energy surface (PES) 
with triaxial quadrupole degrees of freedom. 
The constraints imposed here are on the mass 
quadrupole moments 
that are associated with the polar deformation 
parameters $\beta$ and $\gamma$ 
($0^{\circ}\leqslant\gamma\leqslant60^{\circ}$) \cite{BM}. 
Two types of the mean-field methods are considered: 
(i) the Hartree-Fock-Bogoliubov method 
\cite{robledo2019} with 
the parametrization D1M \cite{D1M} of the 
Gogny EDF \cite{Gogny} for 
$^{132}$Ba and $^{128}$Xe; 
(ii) the relativistic 
Hartree-Bogoliubov method \cite{vretenar2005} 
with the density-dependent point-coupling 
(DD-PC1) EDF \cite{DDPC1} for particle-hole 
channel and the separable pairing force 
of finite range \cite{tian2009} 
for the particle-particle channel 
for $^{134}$Ce and $^{106}$Pd. 
The calculated PESs
for the even-even nuclei 
$^{134}$Ce, $^{132}$Ba, $^{128}$Xe, 
and $^{106}$Pd, shown in Fig.~\ref{fig:pes}, 
are essentially soft in 
$\gamma$ deformation. 
This situation is characteristic of 
the $\gamma$-unstable rotor picture 
\cite{gsoft}, 
which is also equivalent to the O(6) 
limit of the IBM. 

In the next step we build the IBM Hamiltonian 
$\hat H_{\text{B}}$. 
In this study, we employ the proton-neutron 
IBM (IBM-2) \cite{OAI}. The IBM-2 comprises the proton 
$s_{\pi}$ and $d_{\pi}$ bosons, 
and the neutron $s_{\nu}$ 
and $d_{\nu}$ bosons, which represent the 
collective monopole and quadrupole 
proton-proton and neutron-neutron pairs, respectively. 
For the IBM Hamiltonian $\hat H_{\text{B}}$ 
we adopt the form
\begin{align}
\label{eq:ibm2}
 \hat H_{\text{B}} 
= \epsilon_{d}(\hat n_{d_\pi} + \hat n_{d_\nu})
+\kappa\hat{Q}_{\pi}\cdot\hat{Q}_{\nu}, 
\end{align}
where in the first term 
$\hat{n}_{d_\rho}=d^{\dagger}_{\rho}\cdot\tilde{d}_{\rho}$ 
($\rho=\pi$ or $\nu$), 
represents the number operator for the $d_\rho$ bosons, 
with $\epsilon_{d}$ the single $d$-boson energy 
relative to the $s$-boson one, and 
$\tilde{d}_{\mu}=(-1)^{\mu}d_{-\mu}$. 
$\hat Q_{\rho}=s^{\dagger}_{\rho}\tilde
d_{\rho}+d^{\dagger}_{\rho}\tilde
s_{\rho}+\chi_{\rho}(d^{\dagger}_{\rho}\times\tilde{d}_{\rho})^{(2)}$ 
is the bosonic quadrupole operator. 
$\epsilon_{d}$, $\kappa$, $\chi_{\pi}$, 
and $\chi_{\nu}$ are the parameters to be 
determined.

The geometrical structure of a given IBM Hamiltonian 
is studied by introducing the boson 
coherent state \cite{ginocchio1980}, which 
is given by
\begin{align}
\label{eq:coherent}
 \ket{\Phi}=\prod_{\rho=\nu,\pi}
\left[
s_{\rho}^{\+}+\sum_{\mu=-2}^{+2}\alpha_{\rho\mu}d_{\rho\mu}^{\+}
\right]^{N_{\rho}}\ket{0},
\end{align}
up to a normalization factor. 
The amplitudes $\alpha_{\rho\mu}$ are given 
as $\alpha_{\rho0}=\beta_{\rho}\cos{\gamma_{\rho}}$, 
$\alpha_{\rho\pm1}=0$, and 
$\alpha_{\rho\pm2}=\beta_{\rho}\sin{\gamma_{\rho}}/\sqrt{2}$, 
where $\beta_{\rho}$ and $\gamma_{\gamma}$ are 
boson analogs of the deformation variables. 
$N_{\rho}$ is the number of neutron 
($\rho=\nu$) or proton ($\rho=\pi$) bosons, and 
$\ket{0}$ represents the boson vacuum, i.e., the inert core. 
We assume that both proton and neutron bosons have 
equal deformations, $\beta_{\pi}=\beta_{\nu}$ and 
$\gamma_{\pi}=\gamma_{\nu}$. 
We could, in general, take the deformations for 
the proton and neutron bosons to be 
different from each other 
\cite{caprio2004,caprio2005}, 
and would then have to treat 
the energy surface in four dimensions both 
in the mean-field and IBM-2 frameworks. 
In practical calculations, however, comparison between 
the fermionic and bosonic PESs in the four dimensional 
spaces would be too complicated. 
To simplify the discussion, 
we here assume equal proton and neutron deformations 
for both fermion and boson systems. 
We further assume that the fermionic and bosonic 
deformations can be related to each other 
in such a way that 
$\beta_{\pi}=\beta_{\nu}\propto\beta$ 
and $\gamma_{\pi}=\gamma_{\nu}\equiv\gamma$ 
\cite{ginocchio1980,nomura2008}.

\begin{figure}[tb]
\begin{center}
\includegraphics[width=\linewidth]{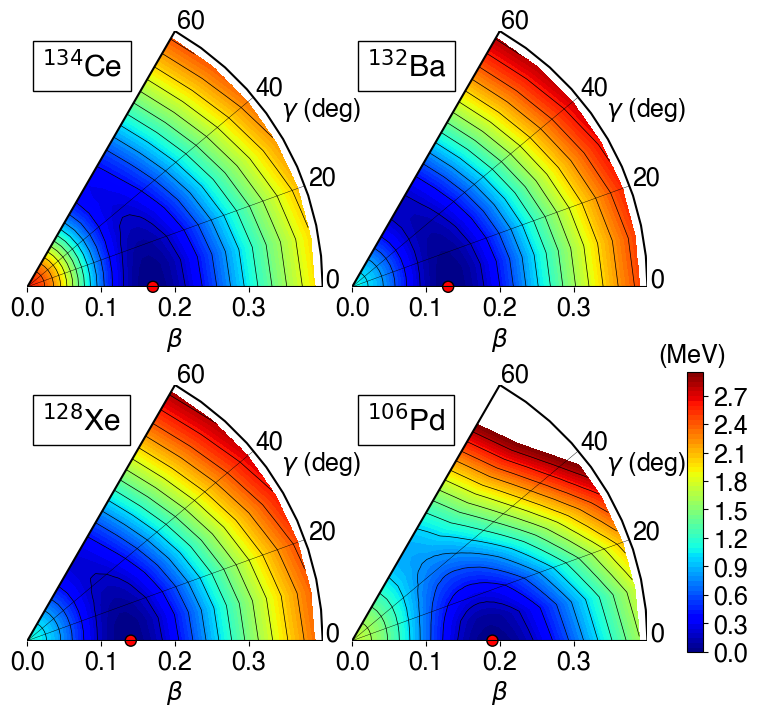}
\caption{Same as Fig.~\ref{fig:pes}, 
but for the bosonic PESs.}
\label{fig:pesibm}
\end{center}
\end{figure}

The parameters of the boson Hamiltonian 
are determined by mapping 
the fermionic PES onto the expectation 
value of $\hat H_{\text{B}}$ in the 
above coherent state, as in Ref.  \cite{nomura2008}. 
In other words, the IBM parameters are 
calibrated 
so that the fermionic and bosonic PESs 
become similar to each other. 
No phenomenological adjustment to experiment 
is made in this procedure. 
We also note that the IBM Hamiltonian (\ref{eq:ibm2}) 
is a rather specific form of the most general IBM-2 
Hamiltonian. A more accurate theoretical description 
of the relevant spectroscopic properties of both even-even 
and odd-mass nuclei might require the inclusion 
of additional terms to 
$\hat H_\text{B}$, in particular, the so-called 
Majorana terms, which could play an important role in 
calculations of $M1$ properties. 
The Majorana terms, however, 
do not add independent contributions to 
the energy surface, unless the proton and neutron 
deformations are taken to be different 
\cite{caprio2005}. 
Under the present assumption that the deformations 
for the proton and 
neutron boson systems are equal to each other, 
the strength parameters of these terms cannot 
be determined 
only by the comparison of the fermionic and 
bosonic PESs that are given in terms of the 
$\beta$ and $\gamma$ degrees of freedom only.

In Fig.~\ref{fig:pesibm} we show 
the mapped (bosonic) PESs for the even-even core nuclei 
$^{134}$Ce, $^{132}$Ba, $^{128}$Xe, and $^{106}$Pd, 
which can be compared with the original 
fermionic PESs in Fig.~\ref{fig:pes}. 
In general, the bosonic PES appears to be 
flat especially in the region far from the 
global minimum. 
This reflects the fact that the IBM is built 
on the valence space of collective nucleon pairs 
in one major shell, 
while the mean-field model involves 
all nucleons. 
The bosonic PESs for $^{134}$Ce,
$^{132}$Ba, and $^{128}$Xe exhibit a 
minimum 
on the prolate axis $\gamma=0^{\circ}$, 
while a shallow triaxial minimum 
at $\gamma\approx20^{\circ}$ is suggested 
in the corresponding fermionic PESs. 
This discrepancy could be corrected by 
including a higher-order term in the 
IBM Hamiltonian \cite{nomura2012tri}. 
However, the triaxial minimum in the fermionic PES 
is so shallow that the discrepancy is expected 
to have a minor influence on the low-lying 
spectra of odd-mass nuclei.

Having fixed the boson-core Hamiltonian, 
we introduce the unpaired nucleon 
degree of freedom. The single-nucleon Hamiltonian 
in Eq.~(\ref{eq:ham}) reads 
\begin{align}
\label{eq:ham-f}
 \hat H_\text{F}
= -\epsilon_{{\jr}}\sqrt{2{\jr}+1}
  (a_{{\jr}}^\dagger\times\tilde{a}_{{\jr}})^{(0)}
\equiv \epsilon_{{\jr}}\hat{n}_{{\jr}},
\end{align}
where $\epsilon_{{\jr}}$ is single-particle energy 
for the odd proton ($\rho=\pi$) or 
neutron ($\rho=\nu$) in orbital $\jr$, 
$a_{{\jr}}^{(\+)}$ denotes the operator 
that annihilate (create) a single nucleon 
with $\tilde{a}_{\jr,m_{\rho}}=(-1)^{{\jr}-m_{\rho}}a_{\jr,-m_{\rho}}$, 
and $\hat{n}_{{\jr}}$ stands for the fermion 
number operator. 

The interaction term 
$\hat V_\mathrm{BF}$ for the 
coupling between the odd nucleon 
with the angular momentum ${\jr}$
and the boson core has the form
\begin{align}
\label{eq:ham-bf}
\hat V_\text{BF}
&=\Gamma_{{\jr}}\hat{Q}_{\rho'}
\cdot(a_{{\jr}}^\dagger\times\tilde{a}_{{\jr}})^{(2)}
+\Lambda_{{\jr}}
\Biggl[
:(s_{\rho'}^\dagger\times\tilde d_{\rho'})^{(2)}
\nonumber\\
&\cdot
\left(
(d_{\rho}^\dagger\times\tilde{a}_{{\jr}})^{(\jr)}
\times
(a_{{\jr}}^\dagger\times\tilde{s}_\rho)^{(\jr)}
\right)^{(2)}:
+ (H.c.)
\Biggr]
\nonumber \\
&+ A_{0}\hat n_{d_{\rho}}\hat{n}_{{\jr}},
\end{align}
where $\rho'\neq\rho$. 
The first, second, and third terms 
are the quadrupole dynamical, exchange, 
and monopole interactions, respectively. 
The $j$-dependent parameters $\Gamma_{\jr}$ 
and $\Lambda_{\jr}$ are given by 
\cite{scholten1985,IBFM}
\begin{subequations}
\label{eq:ham-bf2}
\begin{align}
 \label{eq:ham-bf2a}
 &\Gamma_{{\jr}}=\Gamma_{0}(u^{2}_{{\jr}}-v^{2}_{{\jr}})Q_{{\jr}{\jr}}
\\
\label{eq:ham-bf2b}
&\Lambda_{{\jr}}=\Lambda_{0}
\left[-4u^{2}_{\jr}v^{2}_{{\jr}}Q_{{\jr}{\jr}}^{2}\sqrt{\frac{10}{N_{\rho}(2{\jr}+1)}}\right]
\end{align}
\end{subequations}
$Q_{{\jr}{\jr}}$ stands for the matrix element of the 
spherical harmonic in the single-particle 
basis, i.e., 
$Q_{{\jr}{\jr}}=\braket{l_{\rho}\frac{1}{2}{\jr}\|Y^{(2)}\|l_{\rho}\frac{1}{2}{\jr}}$, 
and $u^{2}_{{\jr}}+v^{2}_{{\jr}}=1$ is satisfied. 
The dots $:(\cdots):$ in Eq.~(\ref{eq:ham-bf}) 
denotes normal ordering. 
Based on the microscopic considerations 
in terms of the generalized seniority scheme 
\cite{scholten1985}, 
it is assumed that both the dynamical 
and exchange terms 
are dominated by the interaction between 
unlike particles, i.e., between
the odd proton (neutron) and the neutron 
(proton) bosons. 
The exchange term takes into account 
 the fact that the bosons are made of 
nucleon pairs. 
For the monopole term, the interaction 
between like-particles, 
i.e., between the odd proton (neutron)
and the proton (neutron) bosons, 
is considered. 
The specific boson-fermion interaction of the form 
(\ref{eq:ham-bf}), which is 
based on the generalized seniority, has been 
frequently used in a number of phenomenological 
IBFM calculations including the spectroscopic 
studies of strongly deformed nuclei 
\cite{IBFM-Book,scholten1985,IBFM}.

The building blocks of $\hat{H}_\mathrm{F}$ 
and $\hat{V}_\mathrm{BF}$ 
are spherical single-particle energy $\epsilon_{\jr}$ 
and occupation probability $v^{2}_{\jr}$ 
of the odd fermion with $j=11/2$, 
which are computed 
by the same mean-field calculations 
constrained to zero deformation \cite{nomura2016odd}. 
The three coupling constants $\Gamma_{0}$, 
$\Lambda_{0}$, and $A_{0}$, defined in 
Eqs.~(\ref{eq:ham-bf}) and (\ref{eq:ham-bf2}), are fitted 
to reproduce to a reasonable accuracy 
the low-lying excitation energies of 
each odd-mass nucleus. 
Table~\ref{tab:para} lists the 
adopted values of the parameters of 
$\hat{H}_{\mathrm{B}}$ (\ref{eq:ibm2}), 
the occupation probability $v^{2}$ of the 
odd particle, and the fitted coupling constants 
of $\hat{V}_{\mathrm{BF}}$. 

The IBFM Hamiltonian thus constructed 
is diagonalized in  
the basis $\ket{[L_{\pi}L_{\nu}(L);j:I}$ 
\cite{PBOS}, 
where $L_{\pi}$ ($L_{\nu}$) and $L$ are the 
angular momentum of proton (neutron) boson system, 
and the total angular momentum of the even-even 
boson core, respectively. 
$I$ stands for the total angular momentum of the
coupled boson-fermion system.

\begin{table}
\caption{\label{tab:para}
The parameters for the IBM Hamiltonian 
(\ref{eq:ibm2}), 
the effective $E2$ boson charge $e^{\mathrm{B}}$, 
proton $g_{\pi}^{\mathrm{B}}$ 
and neutron $g_{\nu}^{\mathrm{B}}$ 
factors, 
and the occupation probability $v^{2}_{\jr}$ of the 
odd particle in the $1h_{11/2}$ single-particle 
state, and the fitted coupling constants 
of $\hat{V}_{\mathrm{BF}}$. 
The numbers of proton $N_{\pi}$ and neutron $N_{\nu}$ 
bosons are also shown, with bar representing hole nature.}
 \begin{center}
 \begin{ruledtabular}
  \begin{tabular}{lcccc}
      & $^{135}$Pr & $^{133}$La & $^{127}$Xe & $^{105}$Pd\\
\hline
Boson core  & $^{134}$Ce & $^{132}$Ba & $^{128}$Xe & $^{106}$Pd\\
$(N_{\pi},N_{\nu})$ & $(4,\bar3)$ & $(3,\bar3)$ & $(2,\bar4)$ & $(\bar2,5)$ \\
$\epsilon_{d}$ (MeV)  & 0.3 & 0.650 & 0.62 & 0.70 \\
$\kappa$ (MeV) & $-0.284$ & $-0.288$ & $-0.315$ & $-0.315$\\
$\chi_{\pi}$ & $-0.45$ & $-0.45$ & $-0.55$ & $-0.45$\\
$\chi_{\nu}$ & 0.25 & 0.25 & 0.25 & $-0.45$\\
$e^{\mathrm{B}}$ ($e\,$b) & 0.12 & 0.123 & 0.10 & 0.095 \\
$g_{\pi}^{\mathrm{B}}$ ($\mu_{N}$) & 1.0 & 1.0 & 1.3 & 1.3 \\
$g_{\nu}^{\mathrm{B}}$ ($\mu_{N}$) & 0.0 & 0.0 & $-0.2$ & 0.3 \\
\hline
Odd particle & $\pi$ & $\pi$ & $\nu$ & $\nu$\\
$v^{2}_{\jr}$ & 0.0303 & 0.03613 & 0.4317 & 0.0494 \\
$\Gamma_{0}$ (MeV) & 0.60 & 0.60 & 1.50 & 1.50 \\
$\Lambda_{0}$ (MeV) & 3.60 & 3.50 & 1.25 & 1.25 \\
$A_{0}$ (MeV) & 0.0 & $-0.5$ & $-0.25$ & $-0.05$ \\
  \end{tabular}
 \end{ruledtabular}
 \end{center}
\end{table}

\section{Results and discussions\label{sec:results}}

\subsection{Excitation spectra}

Now we turn to the discussion about the 
spectroscopic properties of low-spin bands in 
odd-mass nuclei, which are produced by the 
diagonalization of the IBFM Hamiltonian 
(\ref{eq:ham}) with the parameters obtained 
by the aforementioned procedure. 
In what follows, the band 
built on the ${11/2}^{-}_{1}$ state, 
the first, and the second excited bands 
resulting from the IBFM calculations 
are denoted by Yrast, B1, and B2, respectively.

%
%
\begin{figure}[tb]
\begin{center}
\includegraphics[width=\linewidth]{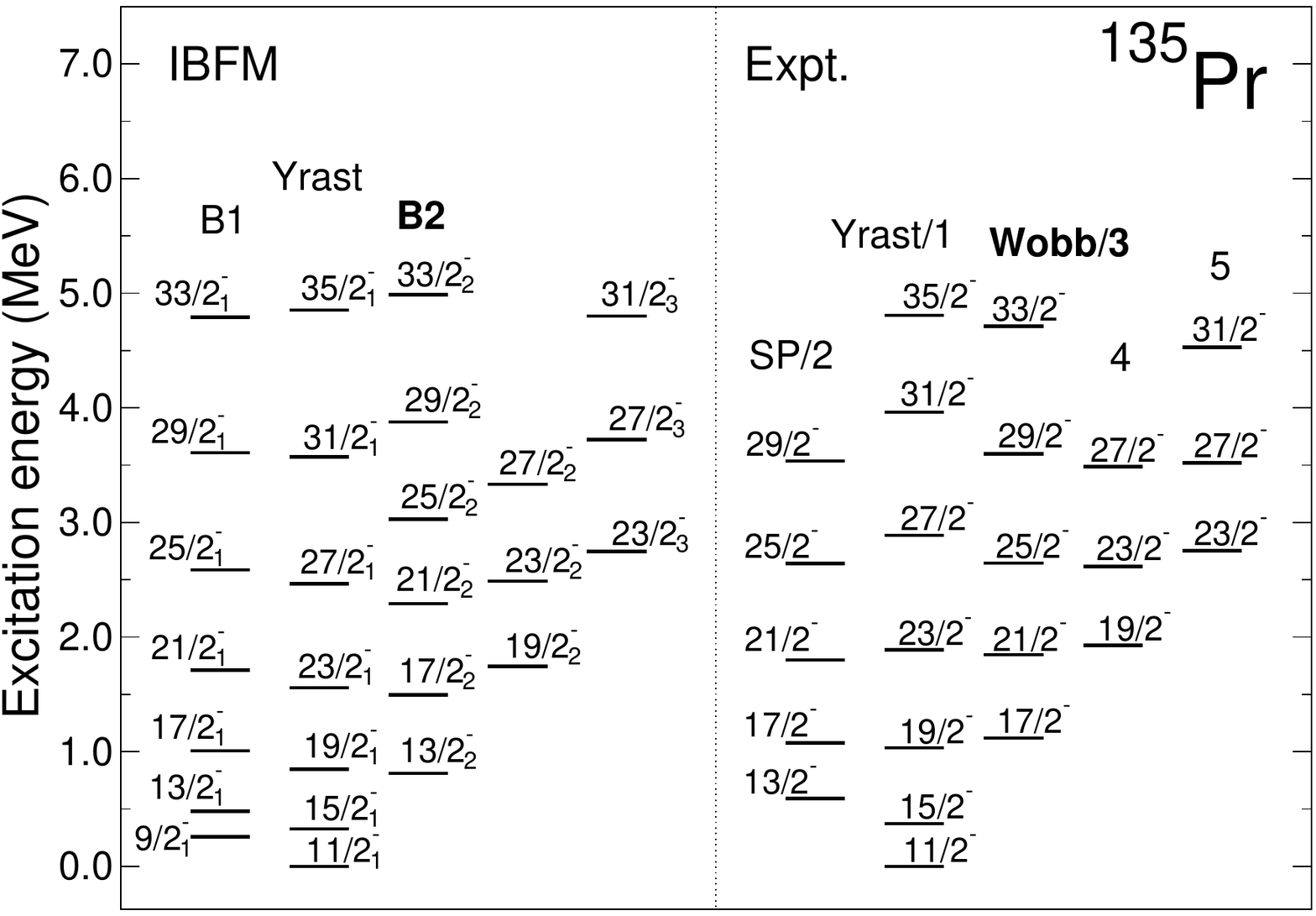}
\caption{Comparisons of theoretical and experimental 
lowest-lying negative-parity 
bands of $^{135}$Pr. The theoretical yrast, 
first, and second excited bands are denoted 
by ``Yrast'', ``B1'', and ``B2'', respectively. 
The experimental bands ``SP'' and ``Wobb'' denote 
the signature-partner and wobbling bands, respectively, 
that were identified in \cite{matta2015}. 
The notations ``1'' to ``5'' for the  
experimental bands 
corresponds to the ones used 
in Lv  {\it et al.}  \cite{135Pr-Lv}. 
The theoretical B2 band should be compared 
with the experimental Wobb band or band 3, 
both of which are highlighted in bold text. 
}
\label{fig:pr135}
\end{center}
\end{figure}

In Fig.~\ref{fig:pr135} 
the predicted five lowest-energy negative-parity 
bands of $^{135}$Pr are shown.  
The theoretical Yrast, B1, and B2 bands 
consist of the states that 
exhibit dominant $\Delta{I}=2$ inband $E2$ 
transitions.
The comparison between the experimental and 
theoretical energy spectra demonstrates that 
the IBFM describes well the 
observed \cite{matta2015,135Pr-Lv} 
low-lying bands in the odd-mass nucleus 
$^{135}$Pr.

For $^{135}$Pr, the experiment performed by 
Matta {\it et al.} \cite{matta2015} showed
that beside the $I={11/2}^{-}$ yrast band, 
the first excited band based on 
the ${13/2}^{-}_{1}$ state 
can be identified as unfavored signature 
partner (SP) band, 
and the second excited band built on 
the ${17/2}^{-}_{2}$ state 
can be assigned as an one-phonon wobbling band 
(denoted by ``Wobb'' in Fig.~\ref{fig:pr135}). 
In the IBFM, band B1 with 
bandhead ${9/2}^{-}_{1}$ 
corresponds to the proposed SP band, 
and band B2 including the ${13/2}^{-}_{2}$, 
${17/2}^{-}_{2}$ $\ldots$ states
is the theoretical counterpart of 
the proposed wobbling band.

As for $^{135}$Pr, 
two additional negative-parity bands 
have been considered in the new measurement 
by Lv  {\it et al.}  \cite{135Pr-Lv}: 
the first one comprising the ${19/2}^{-}_{2}$, 
${23/2}^{-}_{2}$, and ${27/2}^{-}_{2}$ states 
(band 4 in Ref.~\cite{135Pr-Lv} and 
Fig.~\ref{fig:pr135}), 
and the second one comprising the ${23/2}^{-}_{3}$, 
${27/2}^{-}_{3}$, and ${31/2}^{-}_{3}$ states (band 5). 
The IBFM predicts additional two bands built on 
the ${19/2}^{-}_{2}$ and ${23/2}^{-}_{3}$ states, 
which are in agreement with the newly identified 
bands 4 and 5 in Ref. \cite{135Pr-Lv}. 
Our calculation shows that each of the two bands 
is composed by strong inband $\Delta{I}=2$ 
$E2$ transitions. 
In addition, the predicted interband 
$B(E2;{23/2}^{-}_{3}\to{19/2}^{-}_{2})$ transition  
is an order of magnitude weaker than the 
inband $B(E2;{23/2}^{-}_{2}\to{19/2}^{-}_{2})$ 
transition. 
This result supports the 
finding in the new measurement of Ref. \cite{135Pr-Lv}, 
but is in contradiction to the earlier 
experiment of Ref. \cite{sensharma2019}, in which 
the ${19/2}^{-}_{2}$, 
${23/2}^{-}_{3}$, and ${27/2}^{-}_{3}$ 
states were grouped into a single band and interpreted 
as two-phonon wobbling band. 
The new bands proposed in Ref. \cite{135Pr-Lv} 
do not follow the $I(I+1)$ 
energy dependence in a rotor picture, but 
appear to be rather vibrational, a fingerprint 
of the $\gamma$ softness in this mass region.

%
%
\begin{figure}[tb!]
\begin{center}
\includegraphics[width=\linewidth]{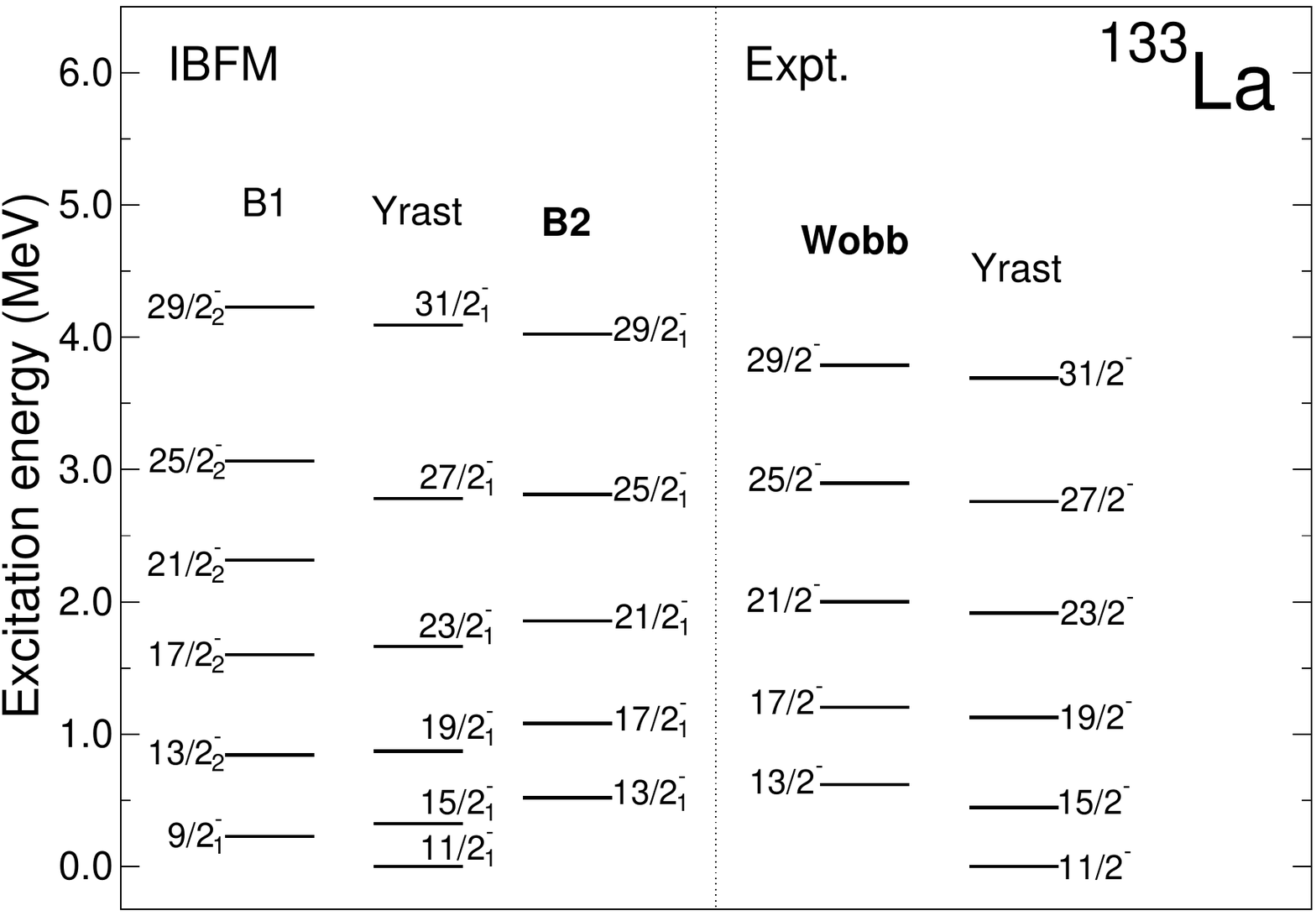}
\caption{Same as Fig.~\ref{fig:pr135}, but for $^{133}$La. 
The notations of the experimental bands 
are according to those used in Ref.~\cite{biswas2019}.}
\label{fig:la133}
\end{center}
\end{figure}

%
%
\begin{figure}[tb!]
\begin{center}
\includegraphics[width=\linewidth]{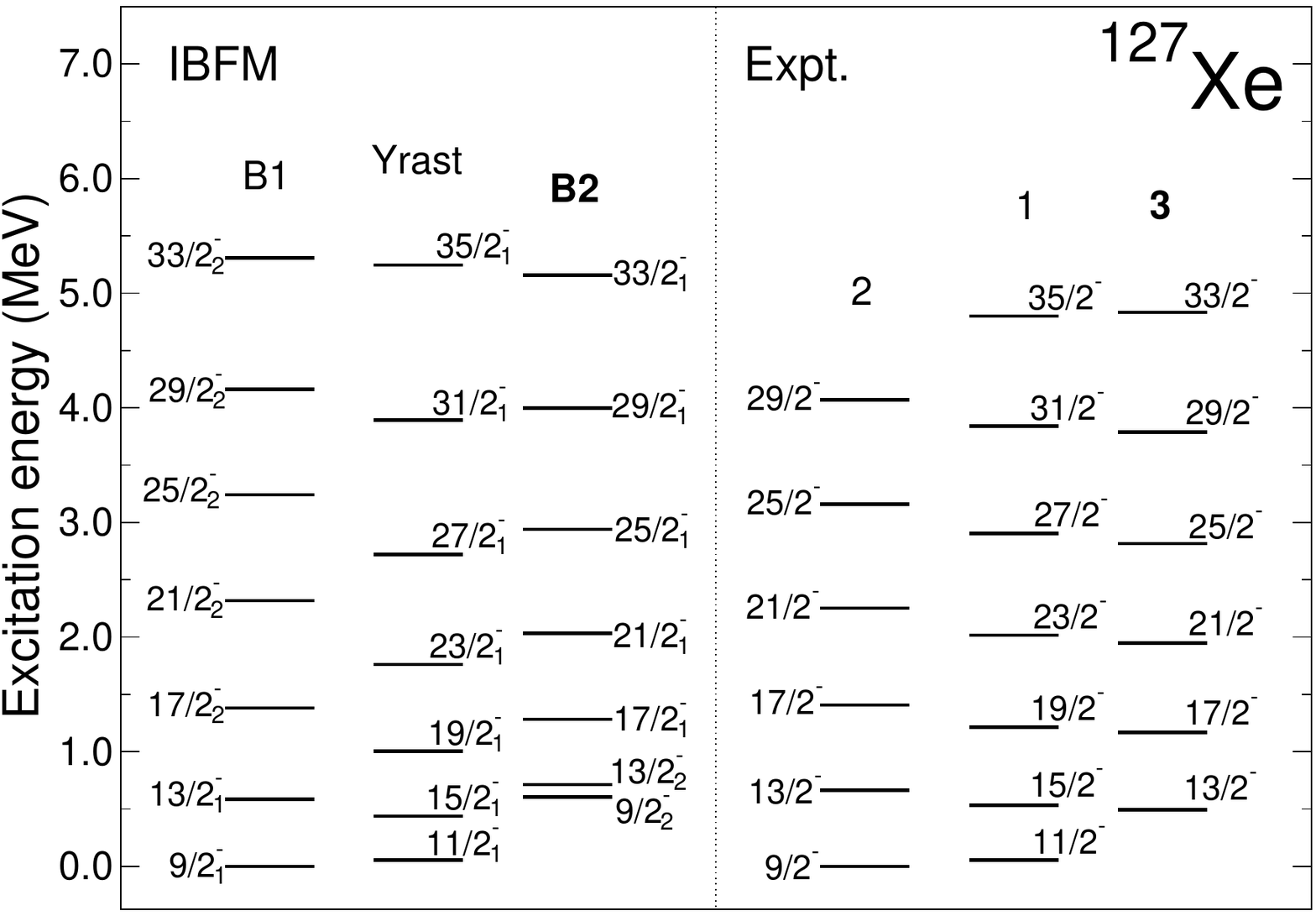}
\caption{Same as Fig.~\ref{fig:pr135}, but for $^{127}$Xe. 
The notations of the experimental bands 
are according to those used in Ref.~\cite{CHAKRABORTY2020}.}
\label{fig:xe127}
\end{center}
\end{figure}

%
%
\begin{figure}[tb!]
\begin{center}
\includegraphics[width=\linewidth]{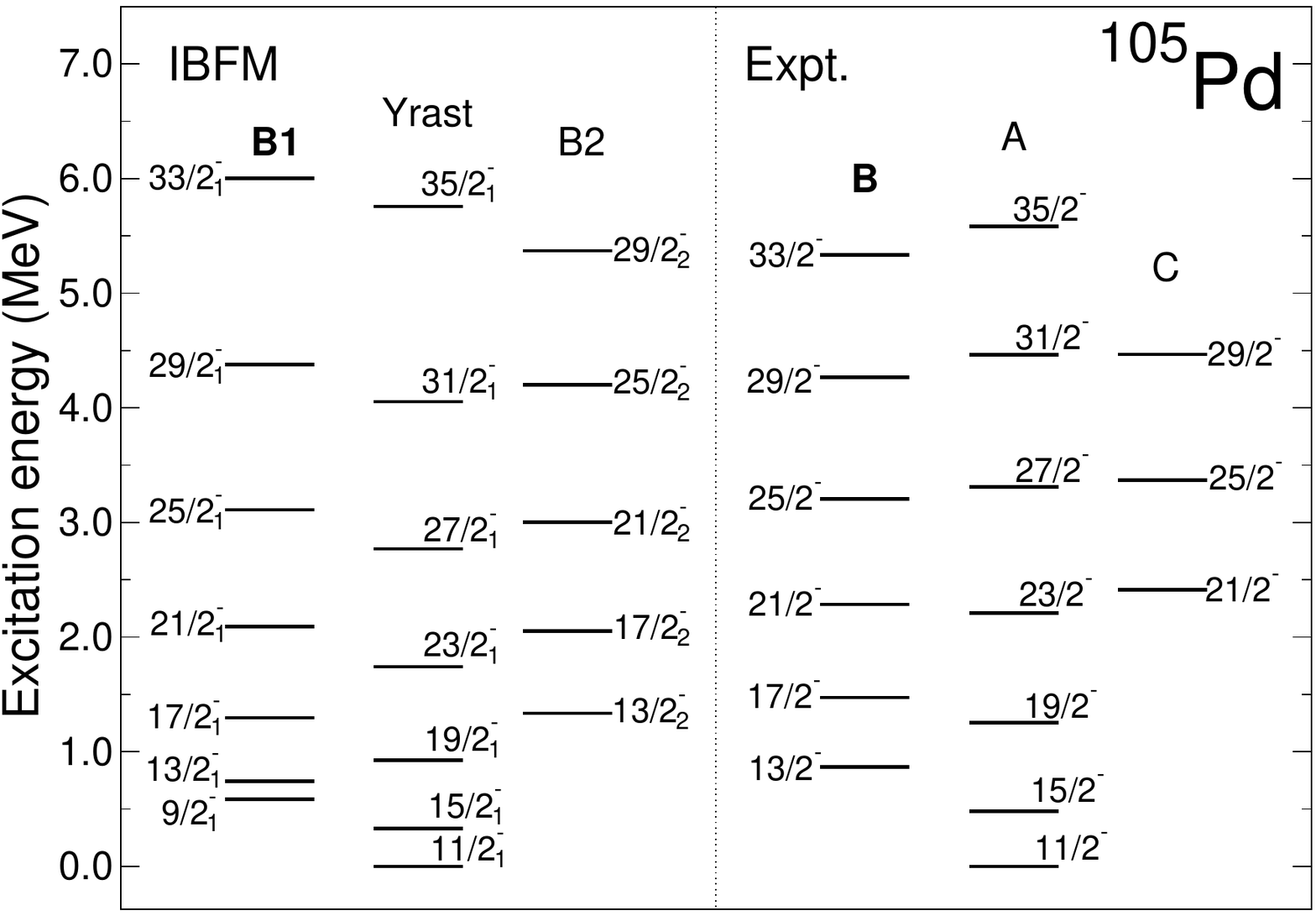}
\caption{Same as Fig.~\ref{fig:pr135}, but for $^{105}$Pd. 
The notations of the experimental bands 
are according to those used in Ref.~\cite{timar2019}.}
\label{fig:pd105}
\end{center}
\end{figure}

Figures~\ref{fig:la133}, \ref{fig:xe127}, 
and \ref{fig:pd105} show the calculated 
and experimental three lowest-lying 
negative-parity bands of the odd-mass nuclei 
$^{133}$La, $^{127}$Xe, and $^{105}$Pd. 
In general, the IBFM description of the observed 
low-lying band structure is satisfactory, 
except for the fact that the spins of the 
predicted bandhead states of some bands 
disagree with the experimental ones. 
The observed bandhead energies of the first (B1) 
and second (B2) excited bands are reproduced well 
by the IBFM. 

For $^{133}$La, besides the ${11/2}^{-}_{1}$ 
ground-state band, the experiment by 
Biswas  {\it et al.}  \cite{biswas2019} 
identified the first excited band based on the 
${13/2}^{-}$ state 
as wobbling band. 
In Fig.~\ref{fig:la133}, we associate the IBFM  
band B2 with bandhead ${13/2}^{-}_{1}$ 
with the proposed wobbling band, according to the 
facts (i) that the calculated energy levels are in a  
better agreement with the experimental ones than 
band B1, (ii) that the spin of the bandhead state 
$I=13/2$ is correctly reproduced, 
and (iii) that, as shown later, the calculated 
$B(E2)_{out}/B(E2)_{in}$ and 
$B(M1)_{out}/B(E2)_{in}$ ratios 
are in a better agreement with 
data \cite{biswas2019} than those for band B1.

The lowest three observed bands for 
$^{127}$Xe \cite{CHAKRABORTY2020} are 
the lowest band (band 2) built on the 
${9/2}^{-}_{1}$ state, 
the first excited band (band 1) built on 
the $I={11/2}^{-}$ state, 
and the second excited band (band 3) built on the 
${13/2}^{-}_{1}$ state, 
which was assigned to be wobbling band 
\cite{CHAKRABORTY2020}. 
Band B2 in the IBFM 
consisting of the ${9/2}^{-}_{2}$, 
${13/2}^{-}_{2}$, ${17/2}^{-}_{1}$, 
${21/2}^{-}_{1}$, ${25/2}^{-}_{1}$, $\ldots$ 
states 
is the theoretical counterpart of band 3. 
The IBFM yields band B1 with the bandhead 
${9/2}^{-}_{1}$ corresponding to the ground state, consistently 
with the observed band 2.

For $^{105}$Pd, Tim\'ar  {\it et al.}  \cite{timar2019} 
interpreted the first excited $\Delta{I}=2$ band 
(band B), which is built on the ${13/2}^{-}_{1}$ 
state 
as wobbling band. 
This is the first evidence for a low-spin wobbling 
band in the mass $A\approx100$ region. 
The present IBFM calculation yields the equivalent 
$\Delta{I}=2$ band (B1) consisting of 
the ${9/2}^{-}_{1}$, 
${13/2}^{-}_{1}$, 
$\ldots$ states.

\subsection{$E2/M1$ mixing ratio}

%
%
\begin{figure}[tb!]
\begin{center}
\includegraphics[width=\linewidth]{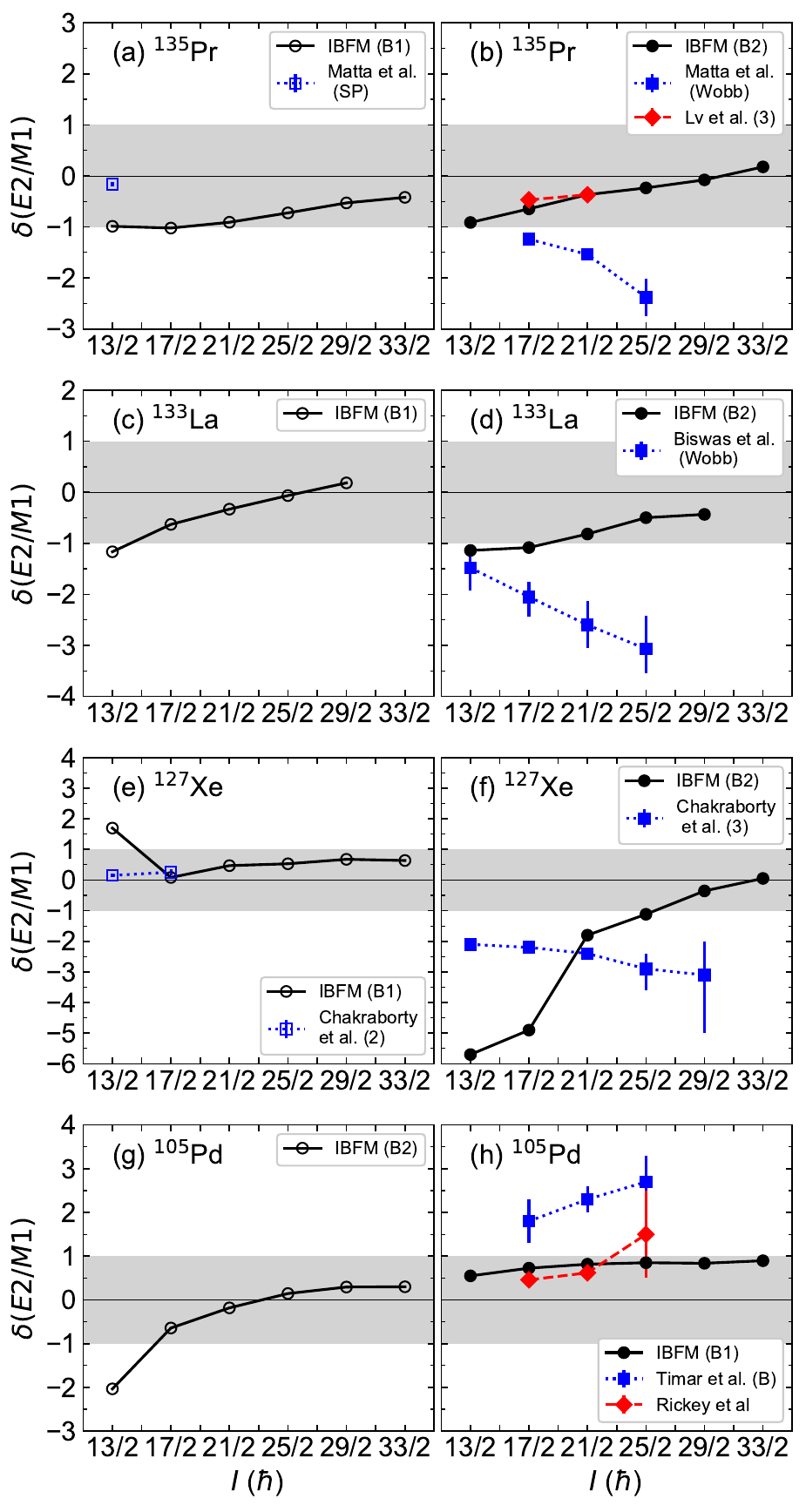}
\caption{The $E2$ to $M1$ mixing ratios $\delta$ 
for the $\Delta{I}=1$ yrare to yrast transitions 
in the odd-mass nuclei 
$^{135}$Pr, $^{133}$La, $^{127}$Xe, 
and $^{105}$Pd, 
for which wobbling bands were 
previously suggested. 
The experimental values 
are taken from 
Refs.~\cite{matta2015,135Pr-Lv,biswas2019,
CHAKRABORTY2020,timar2019,rickey1977}. 
The shaded area in each panel indicates 
$|\delta|<1$. 
The notations of 
the experimental bands in each panel 
follow the ones used in the above references. 
The IBFM $\delta$ values for those bands that were 
previously assigned to be wobbling bands are 
represented by solid 
symbols.}
\label{fig:mratio}
\end{center}
\end{figure}

The $E2$ to $M1$ mixing ratio $\delta$ is 
a criterion for the wobbling interpretation, 
and is calculated by using the formula 
\cite{krane1970,lange1982}
\begin{align}
\label{eq:mratio}
 \delta=0.835\times{E_{\gamma}}
\frac{\braket{I_{f}\|\te\|I_{i}}}
{\braket{I_{f}\|\tm\|I_{i}}},
\end{align}
where $E_{\gamma}=E_{I_{i}}-E_{I_{f}}$, 
with the resultant excitation energies of the 
initial ${I_{i}}$ and final ${I_{f}}$ 
states, and 
${\braket{I_{f} \| \te \| I_{i}}}$
and 
${\braket{I_{f} \| \tm \| I_{i}}}$
represent reduced matrix elements of the $E2$ and 
$M1$ transition operators, respectively. 
The $E2$ operator $\hat T^{E2}$ here takes the form 
\cite{IBFM}:
\begin{align}
 \label{eq:e2}
\te = \teb + \tef,
\end{align}
where
\begin{align}
\teb=e_{\pi}^\mathrm{B}\hat Q_{\pi}
+e_{\nu}^\mathrm{B}\hat Q_{\nu},
\end{align}
and
\begin{align}
\tef=-e^\mathrm{F}
\frac{1}{\sqrt{5}}
(u^{2}_{\jr}-v^{2}_{\jr})Q_{\jr\jr}
(a^{\+}_{\jr}\times\tilde{a}_{\jr})^{(2)}
\end{align}
are the bosonic and fermionic parts of the 
$E2$ operators, respectively. 
Note that $\hat{Q}_{\rho}$ has been defined 
in (\ref{eq:ibm2}). 
We assume that the effective $E2$ charges 
for proton $e^\mathrm{B}_{\pi}$ 
and neutron $e^\mathrm{B}_{\nu}$ bosons 
are equal to each other, 
$e^\mathrm{B}_{\pi} = e^\mathrm{B}_{\nu} \equiv e^{\mathrm{B}}$, 
and fix $e^{\mathrm{B}}$ so that 
the experimental $B(E2; 2^+_1\rightarrow 0^+_1)$ 
rate of each even-even core nucleus is reproduced. 
For the fermion part, standard effective charge 
$e^\mathrm{F}=1.5$ (0.5) 
$e$b is adopted for the odd proton (neutron). 
The $M1$ transition operator $\tm$ reads: 
\begin{align}
 \label{eq:m1}
\tm =\tmb + \tmf
\end{align}
where
\begin{align}
\tmb=\sqrt{\frac{3}{4\pi}}
\left(g_{\pi}^\mathrm{B}\hat L_\pi
+g_{\nu}^\mathrm{B}\hat L_\nu\right), 
\end{align}
and
\begin{align}
\tmf=-\frac{1}{\sqrt{4\pi}}
\braket{j\| g_{l}{\bf l}+g_{s}{\bf s} \|j}
(a_{\jr}^\dagger\times\tilde{a}_{\jr})^{(1)}
\end{align}
are the boson and fermion parts of $\tm$, respectively. 
The effective gyromagnetic ($g$-)factors 
for the proton $g_{\pi}^\mathrm{B}$ 
and neutron $g_\nu^\mathrm{B}$ bosons are
chosen to be close to the empirical values 
\cite{sambataro1984,yoshida2013} that satisfy 
$g_\pi^\mathrm{B}\approx 1.0\,\mu_N$, 
and $g_\nu^\mathrm{B}\approx 0\,\mu_N$. 
For the odd proton (neutron) $g$-factors, 
the standard Schmidt values 
$g_l=1.0\,\mu_N$ and $g_s=5.58\,\mu_N$ 
($g_l=0\,\mu_{N}$ and $g_s=-3.82\,\mu_{N}$)
are used, with $g_s$ quenched by 30\% 
with respect to the free value. 
The adopted values of the boson effective $E2$ 
charge $e^{\mathrm{B}}$, $g$-factors for 
proton $g_{\pi}^{\mathrm{B}}$ 
and neutron $g_{\nu}^{\mathrm{B}}$ 
are found in Table~\ref{tab:para}.

We show in Fig.~\ref{fig:mratio} the calculated 
$\delta(E2/M1)$ ratios for the $\Delta{I}=1$ 
transitions between the yrast and yrare bands. 
The predicted $\delta$ values for $^{135}$Pr 
shown in Figs.~\ref{fig:mratio}(a) and 
\ref{fig:mratio}(b), 
are close to zero 
for both the B1 $\to$ Yrast and B2 $\to$ Yrast 
transitions. 
The experimental absolute 
values $|\delta|$  for the 
Wobb $\to$ Yrast transitions \cite{matta2015} 
increase with spin $I$. 
The updated data of Lv  {\it et al.}  \cite{135Pr-Lv}, 
however, provide smaller mixing ratios 
for the same band (band 3) 
at $I={17/2}$ and ${21/2}$, which agree 
with the present calculations.

The computed mixing ratios for $^{133}$La, shown in 
Figs.~\ref{fig:mratio}(c) and \ref{fig:mratio}(d), 
are generally small, $|\delta|<1$. 
The measured $\delta$ ratios for the 
wobbling (Wobb) band, 
which are the basis of the 
wobbling interpretation of the ${13/2}^{-}$  
band, are depicted in Fig.~\ref{fig:mratio}(d). 
The corresponding IBFM $\delta$ values 
for the B2 $\to$ Yrast transitions, plotted 
also in Fig.~\ref{fig:mratio}(d), are much smaller 
in magnitude than these experimental values, 
but are rather close to 
those obtained with 
the quasiparticle-plus-triaxial-rotor (QTR) 
model \cite{biswas2019}, 
giving more weight to a non-wobbling description 
of the band. We note that the interpretation of the 
experimental ${13/2}^-_1$ band as wobbling band 
\cite{biswas2019} has been recently questioned 
in Ref.~\cite{hua2020}, with respect to the reported 
E2 dominance of the transitions connecting the 
proposed wobbling and normal bands.

The calculated $\delta$ 
for the B1 $\to$ Yrast transitions for $^{127}$Xe 
shown in Fig.~\ref{fig:mratio}(e) are 
small, except for the one at $I={13/2}$. 
As seen in Fig.~\ref{fig:mratio}(f), 
the absolute $\delta$ ratios
for the B2 $\to$ Yrast transitions 
are predicted to 
be smaller in magnitude than the experimental 
counterparts for $I\geqslant {21/2}$. 
However, the predicted ratios are 
unusually large $|\delta|\approx5$ 
for $I<{21/2}$. 
An earlier in-beam spectroscopic study by 
Urban {\it et al.} \cite{urban1984} gave the mixing ratio 
$\delta=-1.7^{+0.4}_{-0.6}$ or $-0.45\pm0.12$ 
for the $E_\gamma=483$ keV (or $13/2^-_1\to11/2^-_1$) 
decay for $^{127}$Xe, the absolute value of which 
is considerably overestimated by the present calculation. 
As we show later, the too large 
$|\delta|$ is here obtained because the calculated 
$M1$ matrix elements for $^{127}$Xe, especially the one 
for the $13/2^-_2\to11/2^-_1$ transition, 
are negligibly small. 
For $^{127}$Xe, the $1h_{11/2}$ single-neutron orbital 
is nearly half filled, $v^{2}_{j_{\nu}}\approx0.5$ 
(see Table~\ref{tab:para}), 
in which case the contributions from both the dynamical 
and exchange terms of $\hat V_\mathrm{BF}$ are rather 
sensitive to the choice of their strength parameters. 
The chosen set of the strength parameters might have yield 
substantial amount of configuration mixing in the lower spin states, 
thus resulting in the too small $M1$ matrix elements.

The absolute values of the calculated 
$\delta$ for both bands B1 and B2 
for $^{105}$Pd shown in Figs.~\ref{fig:mratio}(g) 
and \ref{fig:mratio}(h), 
are all less than one except for the lowest $I={13/2}$ state
of band B2. 
The calculated $\delta$ values 
for band B1 are by a factor of two to three 
smaller than the measured values 
for the proposed wobbling band B \cite{timar2019}. 
In contrast, the mixing ratios obtained by 
Rickey {\it et al.}  \cite{rickey1977}, 
also shown in Fig.~\ref{fig:mratio}(h), 
agree with the calculated values 
at $I={17/2}$ and ${21/2}$. 
In this case we are again faced with contradicting 
experimental values, like in the case of $^{135}$Pr, 
which keeps open the question of the real nature 
of the band.

\subsection{$B(E2)$ and $B(M1)$ transitions}

%
%
\begin{figure}[tb]
\begin{center}
\includegraphics[width=\linewidth]{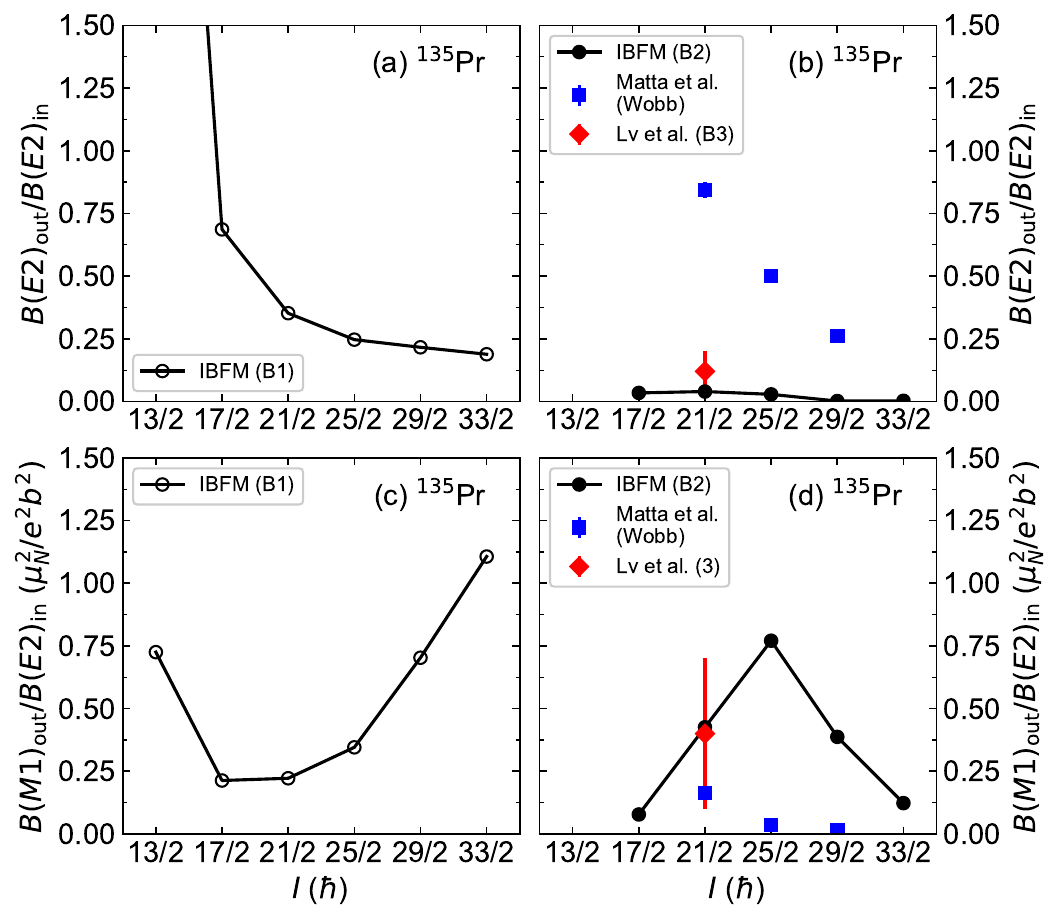}
\caption{The calculated ratios 
$B(E2;I\to I-1)_{out}/B(E2;I\to I-2)_{in}$, and
 $B(M1;I\to I-1)_{out}/B(E2;I\to I-2)_{in}$ 
as functions of spin 
$I$ for the B1 $\to$ Yrast and 
B2 $\to$ Yrast transitions of $^{135}$Pr. 
The notations of the experimental bands 
correspond to those used in 
Refs.~\cite{matta2015,135Pr-Lv}. 
The experimental band built on the 
${17/2}^{-}_{2}$ state was identified as 
wobbling band in \cite{matta2015}, and is 
denoted here by ``Wobb''. 
}
\label{fig:pr135_em}
\end{center}
\end{figure}

In Fig.~\ref{fig:pr135_em} we show the predicted 
$B(E2)_{out}/B(E2)_{in}$ 
($B(M1)_{out}/B(E2)_{in}$) ratios of the interband 
$\Delta{I}=1$ $E2$ ($M1$) to the 
inband $\Delta{I}=2$ $E2$ transitions for  
the bands B1 and B2 for $^{135}$Pr. 
We see from Fig.~\ref{fig:pr135_em}(b)
that, as compared to 
the B1 $\to$ Yrast $E2$ transitions, 
the predicted B2 $\to$ Yrast $E2$ transitions 
are generally weak. 
The calculated $B(E2)_{out}/B(E2)_{in}$ ratios 
for band B2 are much smaller than 
the experimental data reported by Matta {\it et al.}   
\cite{matta2015}, but are closer to the 
new data of Lv  {\it et al.} \cite{135Pr-Lv} at $I={21/2}$. 
The calculated $B(M1)_{out}/B(E2)_{in}$ 
ratios for band B2, 
shown in Fig.~\ref{fig:pr135_em}(d), 
are much larger than the 
data reported in \cite{matta2015}, 
but are in a better agreement with 
the new value at $I={21/2}$ \cite{135Pr-Lv}. 

\begin{figure}[tb]
\begin{center}
\includegraphics[width=\linewidth]{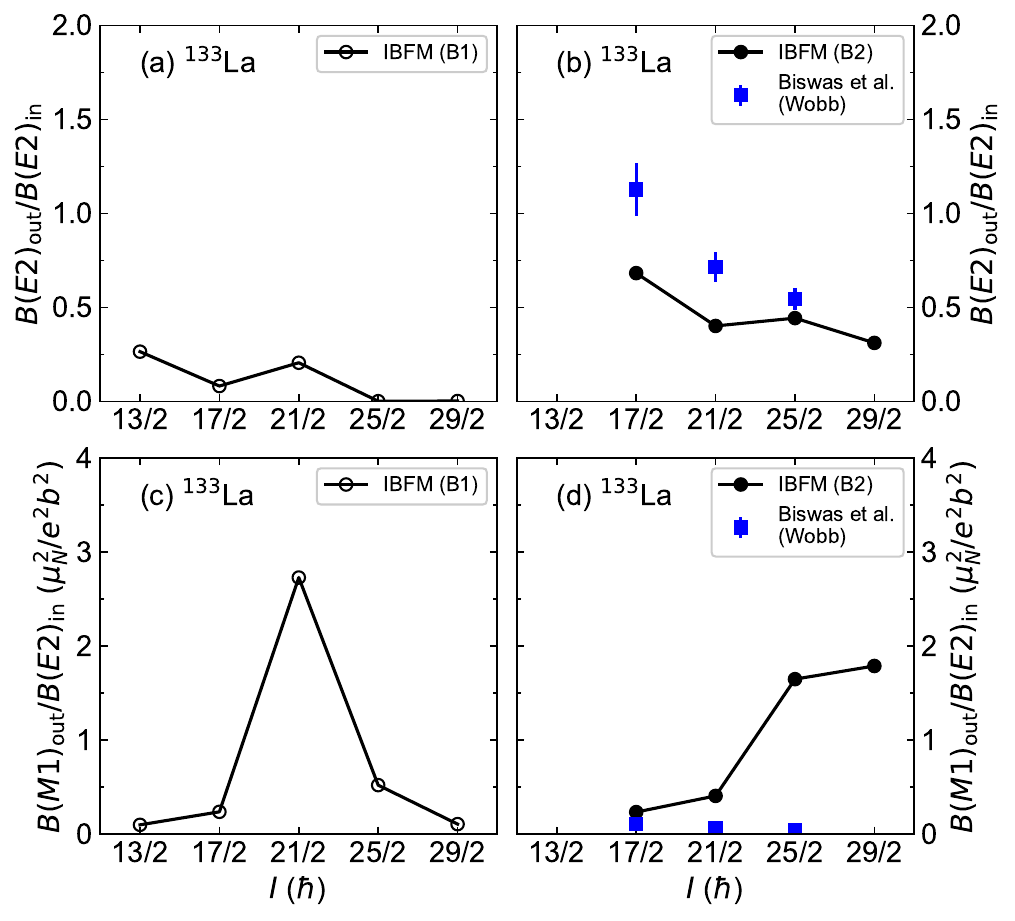}
\caption{Same as Fig.~\ref{fig:pr135_em}, 
but for $^{133}$La. The experimental data 
are from \cite{biswas2019}, which identified 
the yrare band as wobbling band 
(denoted by ``Wobb'').}
\label{fig:la133_em}
\end{center}
\end{figure}

The ratios $B(E2)_{out}/B(E2)_{in}$ 
and $B(M1)_{out}/B(E2)_{in}$ 
for $^{133}$La are shown in Fig.~\ref{fig:la133_em}. 
The predicted $B(E2)_{out}/B(E2)_{in}$ 
ratios for the B2 $\to$ Yrast transitions 
agree with the experimental values reported in 
Ref. \cite{biswas2019} (Fig.~\ref{fig:la133_em}(b)), in which
the first excited band (denoted as ``Wobb'' in 
Figs.~\ref{fig:la133} and \ref{fig:la133_em}) 
has been interpreted as wobbling band, as well as with 
the QTR model calculations \cite{biswas2019}. 
The $B(M1)_{out}/B(E2)_{in}$ 
ratios for the B2 $\to$ Yrast transition are 
calculated to be much larger than the 
measured values of the Wobb $\to$ Yrast 
transitions (Fig.~\ref{fig:la133_em}(d)), 
but rather agree with the QTR results 
\cite{biswas2019}, 
which also overestimated the data.

\begin{figure}[tb]
\begin{center}
\includegraphics[width=\linewidth]{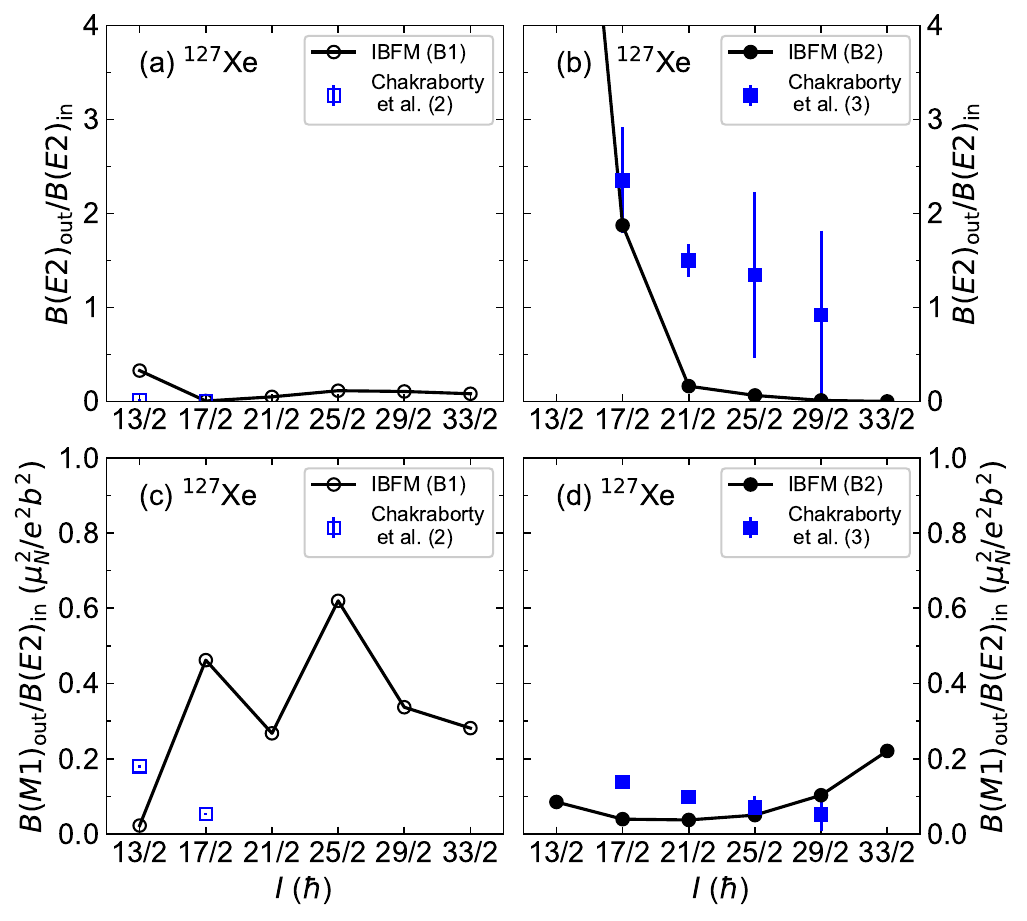}
\caption{Same as Fig.~\ref{fig:la133_em}, 
but for $^{127}$Xe. The experimental data 
are from \cite{CHAKRABORTY2020}, which identified 
the $I={13/2}^{-}$ band (3) as wobbling band.}
\label{fig:xe127_em}
\end{center}
\end{figure}

As for $^{127}$Xe, the calculated 
$B(E2)_{out}/B(E2)_{in}$, shown in Fig.~\ref{fig:xe127_em}(b), 
are much smaller than the experimental ones 
for band 3, which was 
assigned to be wobbling band \cite{CHAKRABORTY2020}. 
The experimental $B(E2)_{out}/B(E2)_{in}$ ratios 
indicate strong $E2$ transitions from 
band 3 to the yrast band 
(band 1 in Fig.~\ref{fig:xe127}), 
especially for the spin $I\geqslant{21/2}$, 
while the errors are also large. 
In Fig.~\ref{fig:xe127_em}(d), 
the predicted $B(M1)_{out}/B(E2)_{in}$ ratios 
for the band B2 
at $I={17/2}$ and ${21/2}$ are considerably 
smaller than the experimental 
values for band 3, which was 
identified as wobbling band. 
We recall that the too large $\delta$ values 
are calculated at these spins (see Fig.~\ref{fig:mratio}(f)).

\begin{figure}[tb]
\begin{center}
\includegraphics[width=\linewidth]{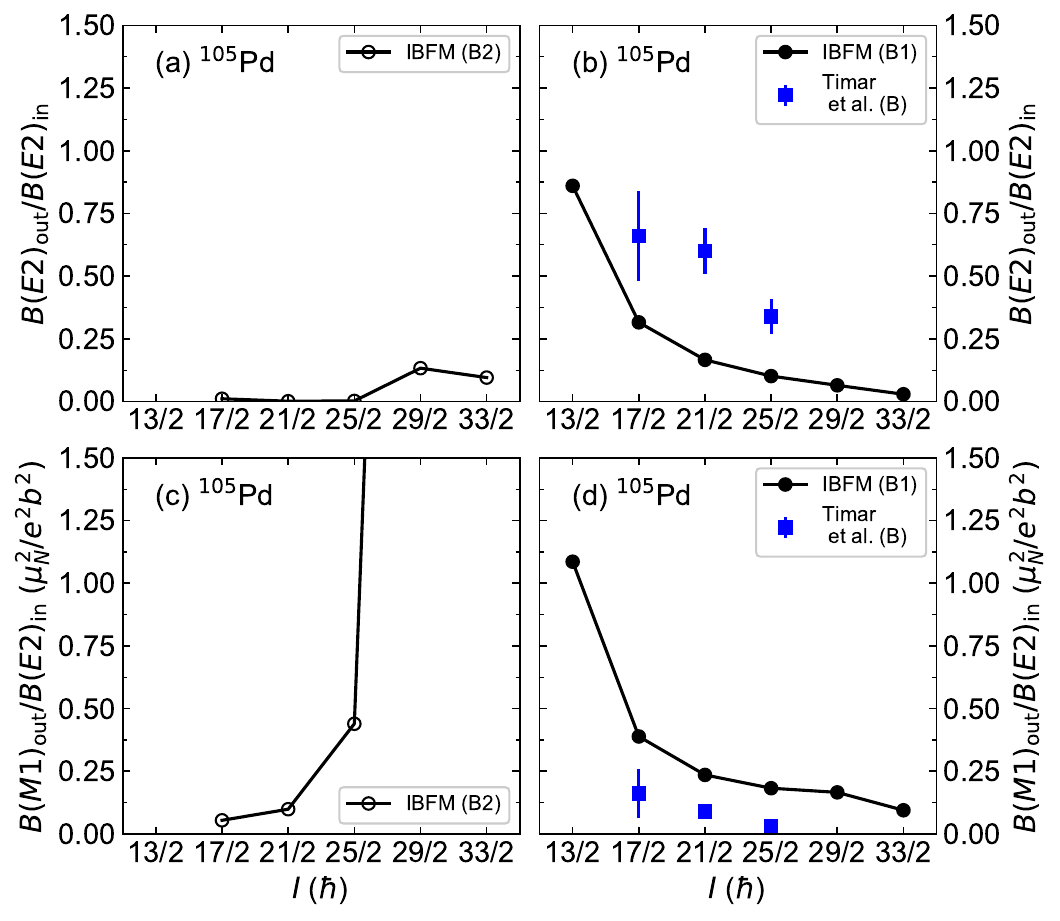}
\caption{Same as Fig.~\ref{fig:la133_em}, 
but for $^{105}$Pd. The experimental band (B), built on 
the ${13/2}^{-}$ state, was identified as wobbling band 
in \cite{timar2019}.}
\label{fig:pd105_em}
\end{center}
\end{figure}

The $B(E2)_{out}/B(E2)_{in}$ 
ratios for both the B1 $\to$ Yrast and 
B2 $\to$ Yrast transitions in $^{105}$Pd shown in Fig.~\ref{fig:pd105_em}
are relatively 
small, in comparison to the measured 
values for 
band B \cite{timar2019}, which was interpreted as 
the wobbling band. 
Generally, the $B(M1)_{out}/B(E2)_{in}$ 
ratios are calculated to be larger 
than the data. 
Thus we can confirm the $M1$ dominance of the 
predicted transitions between yrare and 
yrast bands for $^{105}$Pd.

For the sake of completeness, the calculated 
values for the mixing ratios $\delta$, 
the $B(M1)_{out}/B(E2)_{in}$, the 
$B(E2)_{out}/B(E2)_{in}$ ratios 
of the lowest two excited bands, 
and the corresponding 
experimental values based on the wobbling 
interpretation of the yrare bands 
\cite{matta2015,biswas2019,CHAKRABORTY2020,timar2019},  
and those from the new measurement for $^{135}$Pr 
\cite{135Pr-Lv} and the old experimental $\delta$ 
values for $^{105}$Pd \cite{rickey1977} 
are listed in Table~\ref{tab:em}. 
\begin{table*}[!ht]
 \begin{threeparttable}
  \centering
  \caption{\label{tab:em}
Comparisons between the calculated and experimental 
$\delta(E2/M1)$ mixing ratios of 
the $\Delta I=1$ interband transitions, 
and the ratios of the interband $B(M1;I\to I-1)_{out}$ 
and $B(E2;I\to I-1)_{out}$ to inband 
$B(E2;I\to I-2)_{in}$ transition rates, 
connecting the low-lying yrare bands 
to the yrast bands 
in $^{135}$Pr, $^{133}$La, $^{127}$Xe, and $^{105}$Pd. 
}
 \begin{ruledtabular}
  \begin{tabular}{cccccccccc}
\multirow{2}{*}{Nucleus}
&\multirow{2}{*}{$E_\gamma$ (keV)} &
	   \multirow{2}{*}{Spin} & \multicolumn{2}{c}{$\delta$} &
   \multicolumn{2}{c}{${B(M1)_{out}}/{B(E2)_{in}}$} &
   \multicolumn{2}{c}{${B(E2)_{out}}/{B(E2)_{in}}$} \\
\cline{4-5}
\cline{6-7}
\cline{8-9}
 & & & EXP & IBFM & EXP & IBFM & EXP & IBFM \\
\hline                                        
\multirow{5}{*}{$^{135}$Pr  \cite{matta2015}}
&747.0 &$17/2^-_1$ & $-1.24\pm0.13$ & $-0.646$ & & 0.078 & & 0.034 \\
&812.8 &$21/2^-_1$ & $-1.54\pm0.09$ & $-0.368$ & 0.164$\pm0.014$ & 0.425 & 0.843$\pm0.032$ & 0.040 \\ 
&754.6 &$25/2^-_1$ & $-2.38\pm0.37$ & $-0.236$ & 0.035$\pm0.009$ & 0.771 & 0.500$\pm0.025$ & 0.028 \\ 
&710.2 &$29/2^-_1$ & & $-0.078$ & $\le$0.016$\pm0.004$ & 0.387 & $\ge$0.261$\pm0.014$ & 0.0017 \\ 
&593.9 &$13/2^-_1$ & $-0.16\pm0.04$ & $-0.988$ & & 0.725 & & 4.371 \\
\hline
\multirow{2}{*}{$^{135}$Pr  \cite{135Pr-Lv}}
&747.3 &$17/2^-_1$ & $-0.47_{-0.22}^{+0.09}$ & $-0.646$ & & 0.078 & & 0.034 \\
&813.2 &$21/2^-_1$ & $-0.37_{-0.14}^{+0.10}$ & $-0.368$ & 0.4$\pm0.3$ & 0.425 & 0.12$\pm0.08$ & 0.040 \\                                       
\hline
\multirow{4}{*}{$^{133}$La \cite{biswas2019}}
&618 &$13/2^-_1$ & $-1.48_{-0.32}^{+0.45}$ & $-1.167$ & & & & \\
&758 &$17/2^-_1$ & $-2.05_{-0.30}^{+0.39}$ & $-0.630$ & 0.107$_{-0.028}^{+0.035}$ & 0.232 & 1.127$_{-0.130}^{+0.140}$ & 0.683 \\ 
&874 &$21/2^-_1$ & $-2.60_{-0.47}^{+0.46}$ & $-0.331$ & 0.056$_{-0.019}^{+0.018}$ & 0.404 & 0.716$_{-0.079}^{+0.079}$ & 0.401 \\ 
&982 &$25/2^-_1$ & $-3.07_{-0.65}^{+0.47}$ & $-0.065$ & 0.039$_{-0.015}^{+0.011}$ & 1.646 & 0.545$_{-0.059}^{+0.057}$ & 0.443 \\ 
\hline
\multirow{7}{*}{$^{127}$Xe \cite{CHAKRABORTY2020}}
&483 &$13/2^-_1$ & $-2.1_{-0.2}^{+0.2}$ & $-5.699$ & & 0.085 & & 9.242 \\ 
&639 &$17/2^-_1$ & $-2.2_{-0.1}^{+0.2}$ & $-4.901$ & 0.138$\pm0.012$ & 0.039 & 2.352$\pm0.565$ & 1.874 \\ 
&735 &$21/2^-_1$ & $-2.4_{-0.1}^{+0.1}$ & $-1.801$ & 0.098$\pm0.005$ & 0.037 & 1.500$\pm0.172$ & 0.163 \\
&800 &$25/2^-_1$ & $-2.9_{-0.5}^{+0.7}$ & $-1.117$ & 0.071$\pm0.031$ & 0.050 & 1.346$\pm0.879$ & 0.064 \\ 
&884 &$29/2^-_1$ & $-3.1_{-1.1}^{+1.9}$ & $-0.355$ & 0.052$\pm0.044$ & 0.103 & 0.922$\pm0.895$ & 0.011 \\ 
&651 &$13/2^-_2$ & $+0.15_{-0.05}^{+0.05}$ & $+1.698$ & 0.180$\pm0.004$ & 0.022 & 0.014$\pm0.009$ & 0.329 \\ 
&876 &$17/2^-_2$ & $+0.26_{-0.10}^{+0.10}$ & $+0.085$ & 0.053$\pm0.002$ & 0.462 & 0.007$\pm0.005$ & 0.005 \\
\hline
\multirow{3}{*}{$^{105}$Pd \cite{timar2019}}
&991 &$17/2^-_1$ & $+1.8\pm0.5$ & $+0.727$ & 0.162$\pm0.097$ & 0.316 & 0.66$\pm0.18$ & 0.389 \\ 
&1034&$21/2^-_1$ & $+2.3\pm0.3$ & $+0.817$ & 0.089$\pm0.026$ & 0.166 & 0.60$\pm0.09$ & 0.236 \\
&994 &$25/2^-_1$ & $+2.7\pm0.6$ & $+0.851$ & 0.029$\pm0.057$ & 0.101 & 0.34$\pm0.07$ & 0.182 \\ 
\hline
\multirow{3}{*}{$^{105}$Pd \cite{rickey1977}}
&991 &$17/2^-_1$ & $+0.46\pm0.10$ & $+0.727$ &  & 0.316 &  & 0.389 \\ 
&1034&$21/2^-_1$ & $+0.62\pm0.18$ & $+0.817$ &  & 0.166 &  & 0.236 \\
&994 &$25/2^-_1$ & $+1.5\pm1.0$   & $+0.851$ &  & 0.101 &  & 0.182 \\ 
\end{tabular}
 \end{ruledtabular}
 \end{threeparttable} 
\end{table*}

\begin{figure*}[tb]
\begin{center}
\includegraphics[width=.8\linewidth]{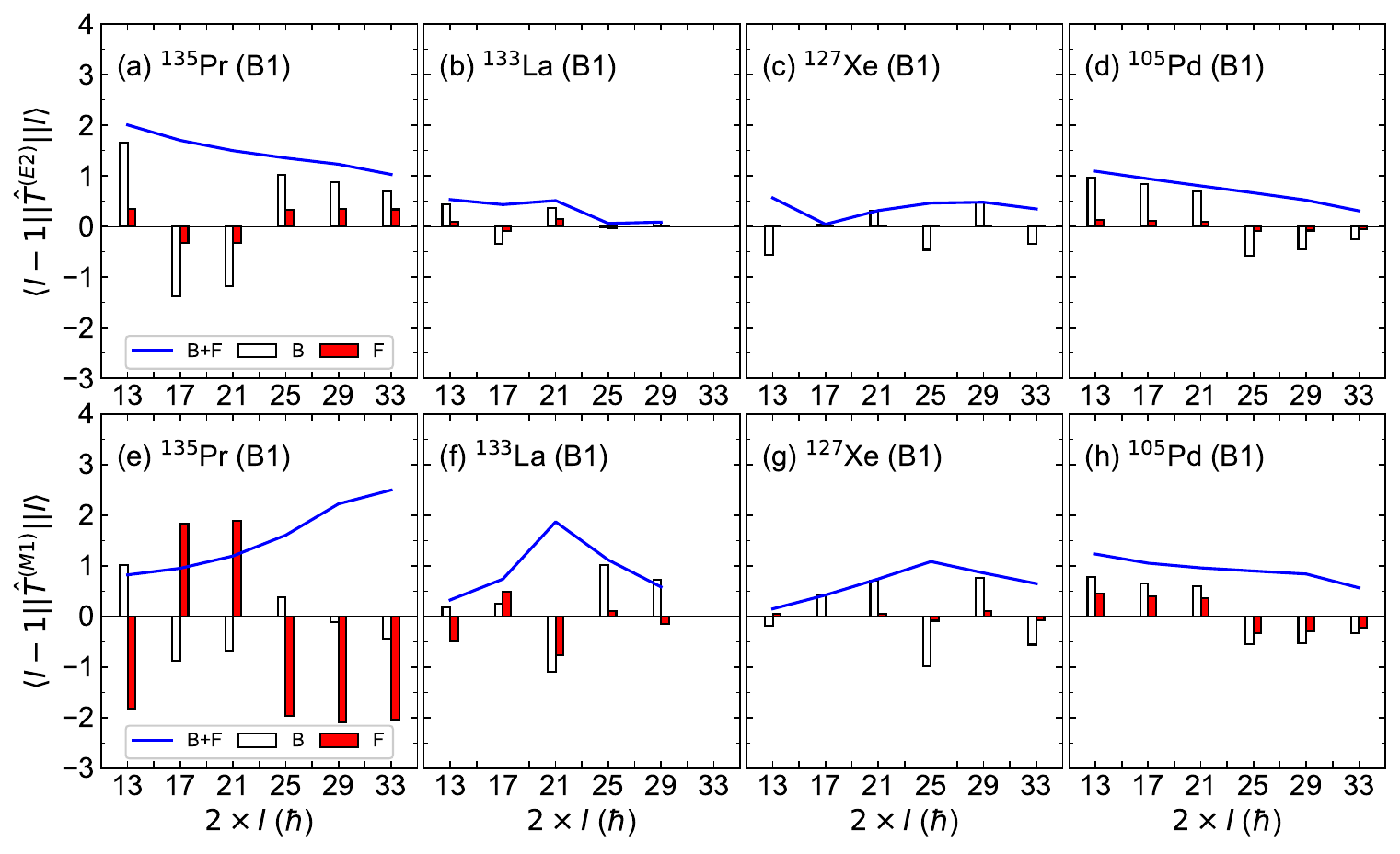}
\caption{The bosonic and fermionic 
reduced matrix elements 
of the $E2$, $\te$ (\ref{eq:e2}), (upper row)
and $M1$, $\tm$ (\ref{eq:m1}), (lower row) operators 
for the $\Delta I=1$ interband transition from 
bands B1 to yrast bands of $^{135}$Pr, 
$^{133}$La, $^{127}$Xe, and $^{105}$Pd, 
plotted for each angular momentum $I$. 
The absolute values of the reduced matrix elements 
of $\te$ and $\tm$ are also shown. 
The theoretical band B1 for $^{105}$Pd is 
here associated with the proposed wobbling 
band in Ref.~\cite{timar2019}.}
\label{fig:rme-b1}
\end{center}
\end{figure*}

\begin{figure*}[tb]
\begin{center}
\includegraphics[width=.8\linewidth]{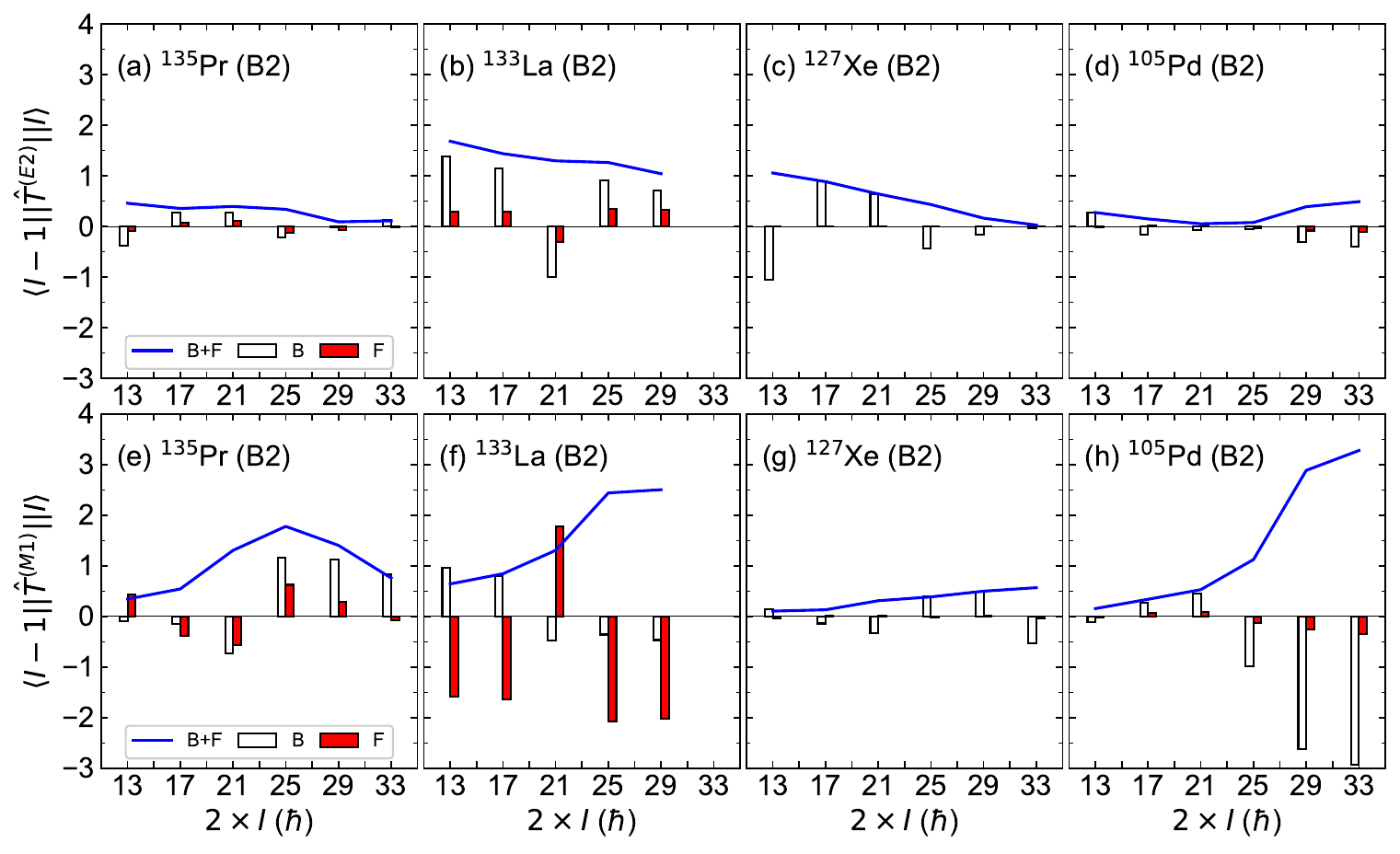}
\caption{The same as Fig.~\ref{fig:rme-b1}, but for the 
transitions from bands B2 to yrast bands. 
The theoretical bands B2 for $^{135}$Pr, 
$^{133}$La, and $^{127}$Xe are here associated 
with the proposed wobbling bands 
in Refs.~\cite{matta2015,biswas2019,CHAKRABORTY2020}, 
respectively.}
\label{fig:rme-b2}
\end{center}
\end{figure*}

\subsection{Boson and fermion contributions to matrix elements}

In this section, we study the individual contributions 
of the boson and fermion parts of the $E2$ (\ref{eq:e2}) 
and $M1$ (\ref{eq:m1}) transition operators 
to the relevant matrix elements. 
In Figs.~\ref{fig:rme-b1} and \ref{fig:rme-b2}, 
we show the corresponding reduced matrix elements of 
the bosonic and fermionic $E2$ ($M1$) operators, 
$\braket{I-1 \| \teb \| I}$ and $\braket{I-1 \| \tef \| I}$
($\braket{I-1 \| \tmb \| I}$ and $\braket{I-1 \| \tmf \| I}$), 
and the absolute values of the full 
matrix elements $|\braket{I-1 \| \te \| I}|$ 
($|\braket{I-1 \| \tm \| I}|$), 
for the $\Delta I=1$ interband transitions 
from the theoretical bands B1 and B2 to yrast 
bands, respectively. 
We observe in these figures that, in general,  
boson contributions are dominant in the $E2$ 
matrix elements, for both bands B1 and B2, and 
for all the nuclei considered. 
In all cases, the boson and fermion parts 
contribute coherently to the $E2$ matrix elements. 
On the other hand, the fermion part plays 
an important role mainly in the $M1$ matrix elements, 
especially, those for band B1 of $^{135}$Pr and band B2 
of $^{133}$La (see Figs.~\ref{fig:rme-b1}(e) and 
\ref{fig:rme-b2}(f)).

For band B1 of $^{135}$Pr 
the fermion contribution to  
the $M1$ matrix elements is systematically 
larger than, and is opposite in sign to the boson one. 
As for band B2 of $^{135}$Pr, which is here associated 
with the proposed wobbling band \cite{matta2015}, 
both the boson and 
fermion parts make small in magnitude, but coherent 
contributions to the $E2$ and $M1$ matrix elements. 
Band B1 (B2) of $^{133}$La appears to show similar 
patterns of the boson and fermion contributions 
to the matrix elements to those of band B2 (B1) 
of $^{135}$Pr. 
We note that band B2 of $^{133}$La is here considered 
to be the theoretical counterpart of the proposed wobbling 
band \cite{biswas2019} (see Fig.~\ref{fig:la133}),  
while both bands B1 and B2 in the same nucleus are shown 
to be close in energy (see Fig.~\ref{fig:la133}) 
and have similar electromagnetic properties 
(see Figs.~\ref{fig:mratio}(c,d) and \ref{fig:la133_em}). 
There is essentially no fermion contribution 
to the $E2$, as well as $M1$, matrix elements of both 
bands B1 and B2 of $^{127}$Xe. For this nucleus, 
the calculated $M1$ 
matrix elements at $I=13/2$ are especially 
small, and this corroborates the too large 
$\delta$ mixing ratios at lower spins 
(see Figs.~\ref{fig:mratio}(e) and \ref{fig:mratio}(f)). 
For band B1 of $^{105}$Pd, 
here associated with the wobbling band \cite{timar2019}, 
the boson and fermion parts make coherent 
contributions to the matrix elements 
(see Figs.~\ref{fig:rme-b1}(d) and \ref{fig:rme-b1}(h)). 
This systematic trend is similar to the one observed for 
band B2 of $^{135}$Pr, which is also associated with 
the wobbling band in the present study. 
The boson contribution to band B2 of $^{105}$Pd 
becomes increasingly larger for higher spin.

\subsection{The gold nuclei}

Finally, we make a remark on the recently 
proposed wobbling bands in the heavier nuclei, 
i.e., $^{187}$Au \cite{sensharma2020} 
and $^{183}$Au \cite{nandi2020}. 
Empirically, their low-lying negative-parity 
states are understood as the proton $\pi{h_{9/2}}$ 
orbital coupled 
with a prolate-deformed core. 
Within the standard IBFM, the above nuclei 
would be described by the coupling between 
the odd proton $\pi h_{11/2}$ orbital 
and the even-even Hg core. 
Earlier phenomenological 
IBFM calculations \cite{bijker1982,arias1986} 
considered the $\pi{h_{9/2}}$ 
intruder orbital coupled to the even-even Pt core 
in order to describe the $I={9/2}^{-}$ ground 
state bands of odd-mass Au nuclei.  
In principle, both the 
$\pi{h_{11/2}}$ and $\pi{h_{9/2}}$
orbitals could be simultaneously included 
in our model but, due to the huge energy 
difference between their 
spherical single-particle levels across the 
proton $Z=82$ major shell gap, 
contribution from the latter is expected 
to be negligible. 
In addition, the even-even core nuclei 
$^{188}$Hg and $^{184}$Hg are often 
characterized by oblate and prolate 
shapes that coexist near the ground state 
\cite{heyde2011}. 
Previous mean-field calculations using the  
Skyrme \cite{yao2013} and Gogny \cite{nomura2013hg} 
forces predicted an oblate minimum 
for $^{188}$Hg and a pronounced prolate-oblate 
shape coexistence for $^{184}$Hg. 
In both cases, the PESs were 
shown to be far from $\gamma$ soft, 
in contrast to the nuclei 
considered here (see Fig.~\ref{fig:pes}). 
These facts indicate a potential difficulty 
in the treatment of the $\pi{h_{11/2}}$ orbital 
for $^{187}$Au and $^{183}$Au 
within our model calculations.

\section{Concluding remarks\label{sec:summary}}

In summary, an alternative interpretation of the 
recently reported low-spin 
wobbling bands in the odd-mass nuclei 
$^{135}$Pr, $^{133}$La, $^{127}$Xe, 
and $^{105}$Pd 
has been presented through 
IBFM calculations. 
The bosonic Hamiltonian for the even-even 
core nuclei, 
and the essential building blocks of the 
boson-fermion interaction 
have been determined by using the 
constrained mean-field approach 
based on a given nuclear EDF. 
The PESs for the even-even core nuclei, 
obtained from the mean-field calculations with 
representative classes of the universal EDF, 
generally exhibit pronounced 
$\gamma$ softness characteristic for nonaxial nuclei. 
The calculated $E2$ to $M1$ 
mixing ratios $\delta$ 
for the $\Delta{I}=1$ transitions between 
the yrare and yrast bands in the considered 
nuclei are consistently small, $|\delta|<1$. 
These mixing ratios indicate the $M1$ 
dominance of the transitions connecting the 
yrare bands in question to the yrast bands, 
which is in contradiction with the wobbling interpretation, 
and are in agreement with the updated 
experimental mixing ratios for $^{135}$Pr \cite{135Pr-Lv} 
and the old data for $^{105}$Pd \cite{rickey1977}. 
This work sheds new light upon the excited 
low-lying bands in $\gamma$-soft nuclei, questioning 
their wobbling interpretation.

\acknowledgements
The authors thank E.~A.~Lawrie for valuable comments. 
The work of K.N.
is supported by the Tenure Track Pilot Programme of 
the Croatian Science Foundation and the 
\'Ecole Polytechnique F\'ed\'erale de Lausanne,
and the Project TTP-2018-07-3554 Exotic Nuclear Structure
and Dynamics, with funds of the Croatian-Swiss Research
Programme. 

\bibliography{refs}

\begin{thebibliography}{61}%
\makeatletter
\providecommand \@ifxundefined [1]{%
 \@ifx{#1\undefined}
}%
\providecommand \@ifnum [1]{%
 \ifnum #1\expandafter \@firstoftwo
 \else \expandafter \@secondoftwo
 \fi
}%
\providecommand \@ifx [1]{%
 \ifx #1\expandafter \@firstoftwo
 \else \expandafter \@secondoftwo
 \fi
}%
\providecommand \natexlab [1]{#1}%
\providecommand \enquote  [1]{``#1''}%
\providecommand \bibnamefont  [1]{#1}%
\providecommand \bibfnamefont [1]{#1}%
\providecommand \citenamefont [1]{#1}%
\providecommand \href@noop [0]{\@secondoftwo}%
\providecommand \href [0]{\begingroup \@sanitize@url \@href}%
\providecommand \@href[1]{\@@startlink{#1}\@@href}%
\providecommand \@@href[1]{\endgroup#1\@@endlink}%
\providecommand \@sanitize@url [0]{\catcode `\\12\catcode `\$12\catcode
  `\&12\catcode `\#12\catcode `\^12\catcode `\_12\catcode `\%12\relax}%
\providecommand \@@startlink[1]{}%
\providecommand \@@endlink[0]{}%
\providecommand \url  [0]{\begingroup\@sanitize@url \@url }%
\providecommand \@url [1]{\endgroup\@href {#1}{\urlprefix }}%
\providecommand \urlprefix  [0]{URL }%
\providecommand \Eprint [0]{\href }%
\providecommand \doibase [0]{https://doi.org/}%
\providecommand \selectlanguage [0]{\@gobble}%
\providecommand \bibinfo  [0]{\@secondoftwo}%
\providecommand \bibfield  [0]{\@secondoftwo}%
\providecommand \translation [1]{[#1]}%
\providecommand \BibitemOpen [0]{}%
\providecommand \bibitemStop [0]{}%
\providecommand \bibitemNoStop [0]{.\EOS\space}%
\providecommand \EOS [0]{\spacefactor3000\relax}%
\providecommand \BibitemShut  [1]{\csname bibitem#1\endcsname}%
\let\auto@bib@innerbib\@empty
\bibitem [{\citenamefont {Bohr}\ and\ \citenamefont
  {Mottelson}(1975{\natexlab{a}})}]{BM_II}%
  \BibitemOpen
  \bibfield  {author} {\bibinfo {author} {\bibfnamefont {A.}~\bibnamefont
  {Bohr}}\ and\ \bibinfo {author} {\bibfnamefont {B.~R.}\ \bibnamefont
  {Mottelson}},\ }\href@noop {} {\emph {\bibinfo {title} {Nuclear
  Structure}}},\ Vol.~\bibinfo {volume} {II}\ (\bibinfo  {publisher} {Benjamin,
  New York, USA},\ \bibinfo {year} {1975})\BibitemShut {NoStop}%
\bibitem [{\citenamefont {\O{}deg\aa{}rd~{\it et
  al.}}(2001)}]{wobblingprl-2001}%
  \BibitemOpen
  \bibfield  {author} {\bibinfo {author} {\bibfnamefont {S.~W.}\ \bibnamefont
  {\O{}deg\aa{}rd~{\it et al.}}},\ }\href
  {https://doi.org/10.1103/PhysRevLett.86.5866} {\bibfield  {journal} {\bibinfo
   {journal} {Phys. Rev. Lett.}\ }\textbf {\bibinfo {volume} {86}},\ \bibinfo
  {pages} {5866} (\bibinfo {year} {2001})}\BibitemShut {NoStop}%
\bibitem [{\citenamefont {Jensen~{\it et al.}}(2002)}]{wobblingprl-2002}%
  \BibitemOpen
  \bibfield  {author} {\bibinfo {author} {\bibfnamefont {D.~R.}\ \bibnamefont
  {Jensen~{\it et al.}}},\ }\href
  {https://doi.org/10.1103/PhysRevLett.89.142503} {\bibfield  {journal}
  {\bibinfo  {journal} {Phys. Rev. Lett.}\ }\textbf {\bibinfo {volume} {89}},\
  \bibinfo {pages} {142503} (\bibinfo {year} {2002})}\BibitemShut {NoStop}%
\bibitem [{\citenamefont {Sch{\"o}nwa{\ss}er~{\it et al.}}(2003)}]{165Lu}%
  \BibitemOpen
  \bibfield  {author} {\bibinfo {author} {\bibfnamefont {G.}~\bibnamefont
  {Sch{\"o}nwa{\ss}er~{\it et al.}}},\ }\href
  {https://doi.org/https://doi.org/10.1016/S0370-2693(02)03095-2} {\bibfield
  {journal} {\bibinfo  {journal} {Phys. Lett. B}\ }\textbf {\bibinfo {volume}
  {552}},\ \bibinfo {pages} {9 } (\bibinfo {year} {2003})}\BibitemShut
  {NoStop}%
\bibitem [{\citenamefont {Amro~{\it et al.}}(2003)}]{167Lu}%
  \BibitemOpen
  \bibfield  {author} {\bibinfo {author} {\bibfnamefont {H.}~\bibnamefont
  {Amro~{\it et al.}}},\ }\href
  {https://doi.org/https://doi.org/10.1016/S0370-2693(02)03199-4} {\bibfield
  {journal} {\bibinfo  {journal} {Phys. Lett. B}\ }\textbf {\bibinfo {volume}
  {553}},\ \bibinfo {pages} {197 } (\bibinfo {year} {2003})}\BibitemShut
  {NoStop}%
\bibitem [{\citenamefont {Bringel~{\it et al.}}(2005)}]{161Lu}%
  \BibitemOpen
  \bibfield  {author} {\bibinfo {author} {\bibfnamefont {P.}~\bibnamefont
  {Bringel~{\it et al.}}},\ }\href {https://doi.org/10.1140/epja/i2005-10005-7}
  {\bibfield  {journal} {\bibinfo  {journal} {Eur. Phys. J. A}\ }\textbf
  {\bibinfo {volume} {24}},\ \bibinfo {pages} {167} (\bibinfo {year}
  {2005})}\BibitemShut {NoStop}%
\bibitem [{\citenamefont {Hartley~{\it et al.}}(2009)}]{167Ta}%
  \BibitemOpen
  \bibfield  {author} {\bibinfo {author} {\bibfnamefont {D.~J.}\ \bibnamefont
  {Hartley~{\it et al.}}},\ }\href {https://doi.org/10.1103/PhysRevC.80.041304}
  {\bibfield  {journal} {\bibinfo  {journal} {Phys. Rev. C}\ }\textbf {\bibinfo
  {volume} {80}},\ \bibinfo {pages} {041304} (\bibinfo {year}
  {2009})}\BibitemShut {NoStop}%
\bibitem [{\citenamefont {Matta~{\it et al.}}(2015)}]{matta2015}%
  \BibitemOpen
  \bibfield  {author} {\bibinfo {author} {\bibfnamefont {J.~T.}\ \bibnamefont
  {Matta~{\it et al.}}},\ }\href
  {https://doi.org/10.1103/PhysRevLett.114.082501} {\bibfield  {journal}
  {\bibinfo  {journal} {Phys. Rev. Lett.}\ }\textbf {\bibinfo {volume} {114}},\
  \bibinfo {pages} {082501} (\bibinfo {year} {2015})}\BibitemShut {NoStop}%
\bibitem [{\citenamefont {{Sensharma {\it et al.}}}(2019)}]{sensharma2019}%
  \BibitemOpen
  \bibfield  {author} {\bibinfo {author} {\bibfnamefont {N.}~\bibnamefont
  {{Sensharma {\it et al.}}}},\ }\href
  {https://doi.org/https://doi.org/10.1016/j.physletb.2019.03.038} {\bibfield
  {journal} {\bibinfo  {journal} {Phys. Lett. B}\ }\textbf {\bibinfo {volume}
  {792}},\ \bibinfo {pages} {170} (\bibinfo {year} {2019})}\BibitemShut
  {NoStop}%
\bibitem [{\citenamefont {Biswas~{\it et al.}}(2019)}]{biswas2019}%
  \BibitemOpen
  \bibfield  {author} {\bibinfo {author} {\bibfnamefont {S.}~\bibnamefont
  {Biswas~{\it et al.}}},\ }\href {https://doi.org/10.1140/epja/i2019-12856-5}
  {\bibfield  {journal} {\bibinfo  {journal} {Eur. Phys. J. A}\ }\textbf
  {\bibinfo {volume} {55}},\ \bibinfo {pages} {159} (\bibinfo {year}
  {2019})}\BibitemShut {NoStop}%
\bibitem [{\citenamefont {Tim\'ar~{\it et al.}}(2019)}]{timar2019}%
  \BibitemOpen
  \bibfield  {author} {\bibinfo {author} {\bibfnamefont {J.}~\bibnamefont
  {Tim\'ar~{\it et al.}}},\ }\href
  {https://doi.org/10.1103/PhysRevLett.122.062501} {\bibfield  {journal}
  {\bibinfo  {journal} {Phys. Rev. Lett.}\ }\textbf {\bibinfo {volume} {122}},\
  \bibinfo {pages} {062501} (\bibinfo {year} {2019})}\BibitemShut {NoStop}%
\bibitem [{\citenamefont {{Chakraborty {\it et al.}}}(2020)}]{CHAKRABORTY2020}%
  \BibitemOpen
  \bibfield  {author} {\bibinfo {author} {\bibfnamefont {S.}~\bibnamefont
  {{Chakraborty {\it et al.}}}},\ }\href
  {https://doi.org/https://doi.org/10.1016/j.physletb.2020.135854} {\bibfield
  {journal} {\bibinfo  {journal} {Phys. Lett. B}\ }\textbf {\bibinfo {volume}
  {811}},\ \bibinfo {pages} {135854} (\bibinfo {year} {2020})}\BibitemShut
  {NoStop}%
\bibitem [{\citenamefont {Sensharma~{\it et al.}}(2020)}]{sensharma2020}%
  \BibitemOpen
  \bibfield  {author} {\bibinfo {author} {\bibfnamefont {N.}~\bibnamefont
  {Sensharma~{\it et al.}}},\ }\href
  {https://doi.org/10.1103/PhysRevLett.124.052501} {\bibfield  {journal}
  {\bibinfo  {journal} {Phys. Rev. Lett.}\ }\textbf {\bibinfo {volume} {124}},\
  \bibinfo {pages} {052501} (\bibinfo {year} {2020})}\BibitemShut {NoStop}%
\bibitem [{\citenamefont {Nandi~{\it et al.}}(2020)}]{nandi2020}%
  \BibitemOpen
  \bibfield  {author} {\bibinfo {author} {\bibfnamefont {S.}~\bibnamefont
  {Nandi~{\it et al.}}},\ }\href
  {https://doi.org/10.1103/PhysRevLett.125.132501} {\bibfield  {journal}
  {\bibinfo  {journal} {Phys. Rev. Lett.}\ }\textbf {\bibinfo {volume} {125}},\
  \bibinfo {pages} {132501} (\bibinfo {year} {2020})}\BibitemShut {NoStop}%
\bibitem [{\citenamefont {Chen}\ \emph {et~al.}(2019)\citenamefont {Chen},
  \citenamefont {Frauendorf},\ and\ \citenamefont {Petrache}}]{chen-2019}%
  \BibitemOpen
  \bibfield  {author} {\bibinfo {author} {\bibfnamefont {Q.~B.}\ \bibnamefont
  {Chen}}, \bibinfo {author} {\bibfnamefont {S.}~\bibnamefont {Frauendorf}},\
  and\ \bibinfo {author} {\bibfnamefont {C.~M.}\ \bibnamefont {Petrache}},\
  }\href {https://doi.org/10.1103/PhysRevC.100.061301} {\bibfield  {journal}
  {\bibinfo  {journal} {Phys. Rev. C}\ }\textbf {\bibinfo {volume} {100}},\
  \bibinfo {pages} {061301(R)} (\bibinfo {year} {2019})}\BibitemShut {NoStop}%
\bibitem [{\citenamefont {Chen}\ and\ \citenamefont
  {Petrache}(2021)}]{chen-2021}%
  \BibitemOpen
  \bibfield  {author} {\bibinfo {author} {\bibfnamefont {F.-Q.}\ \bibnamefont
  {Chen}}\ and\ \bibinfo {author} {\bibfnamefont {C.~M.}\ \bibnamefont
  {Petrache}},\ }\href {https://doi.org/10.1103/PhysRevC.103.064319} {\bibfield
   {journal} {\bibinfo  {journal} {Phys. Rev. C}\ }\textbf {\bibinfo {volume}
  {103}},\ \bibinfo {pages} {064319} (\bibinfo {year} {2021})}\BibitemShut
  {NoStop}%
\bibitem [{\citenamefont {Guo~{\it et al.}}()}]{Guo}%
  \BibitemOpen
  \bibfield  {author} {\bibinfo {author} {\bibfnamefont {S.}~\bibnamefont
  {Guo~{\it et al.}}},\ }\href@noop {} {\bibinfo {title} {to be published}},\
  \bibinfo {howpublished} {{https://arxiv.org/abs/2011.143549}}\BibitemShut
  {NoStop}%
\bibitem [{\citenamefont {Lv~{\it et al.}}(2022)}]{135Pr-Lv}%
  \BibitemOpen
  \bibfield  {author} {\bibinfo {author} {\bibfnamefont {B.~F.}\ \bibnamefont
  {Lv~{\it et al.}}},\ }\href
  {https://doi.org/https://doi.org/10.1016/j.physletb.2021.136840} {\bibfield
  {journal} {\bibinfo  {journal} {Phys. Lett. B}\ }\textbf {\bibinfo {volume}
  {824}},\ \bibinfo {pages} {136840} (\bibinfo {year} {2022})}\BibitemShut
  {NoStop}%
\bibitem [{\citenamefont {Frauendorf}\ and\ \citenamefont
  {D\"onau}(2014)}]{Transverse-2014}%
  \BibitemOpen
  \bibfield  {author} {\bibinfo {author} {\bibfnamefont {S.}~\bibnamefont
  {Frauendorf}}\ and\ \bibinfo {author} {\bibfnamefont {F.}~\bibnamefont
  {D\"onau}},\ }\href {https://doi.org/10.1103/PhysRevC.89.014322} {\bibfield
  {journal} {\bibinfo  {journal} {Phys. Rev. C}\ }\textbf {\bibinfo {volume}
  {89}},\ \bibinfo {pages} {014322} (\bibinfo {year} {2014})}\BibitemShut
  {NoStop}%
\bibitem [{\citenamefont {Iachello}\ and\ \citenamefont
  {Scholten}(1979)}]{iachello1979}%
  \BibitemOpen
  \bibfield  {author} {\bibinfo {author} {\bibfnamefont {F.}~\bibnamefont
  {Iachello}}\ and\ \bibinfo {author} {\bibfnamefont {O.}~\bibnamefont
  {Scholten}},\ }\href@noop {} {\bibfield  {journal} {\bibinfo  {journal}
  {Phys. Rev. Lett.}\ }\textbf {\bibinfo {volume} {43}},\ \bibinfo {pages}
  {679} (\bibinfo {year} {1979})}\BibitemShut {NoStop}%
\bibitem [{\citenamefont {Iachello}\ and\ \citenamefont {{Van
  Isacker}}(1991)}]{IBFM}%
  \BibitemOpen
  \bibfield  {author} {\bibinfo {author} {\bibfnamefont {F.}~\bibnamefont
  {Iachello}}\ and\ \bibinfo {author} {\bibfnamefont {P.}~\bibnamefont {{Van
  Isacker}}},\ }\href@noop {} {\emph {\bibinfo {title} {The interacting
  boson-fermion model}}}\ (\bibinfo  {publisher} {Cambridge University Press,
  Cambridge},\ \bibinfo {year} {1991})\BibitemShut {NoStop}%
\bibitem [{\citenamefont {Bender}\ \emph {et~al.}(2003)\citenamefont {Bender},
  \citenamefont {Heenen},\ and\ \citenamefont {Reinhard}}]{bender2003}%
  \BibitemOpen
  \bibfield  {author} {\bibinfo {author} {\bibfnamefont {M.}~\bibnamefont
  {Bender}}, \bibinfo {author} {\bibfnamefont {P.-H.}\ \bibnamefont {Heenen}},\
  and\ \bibinfo {author} {\bibfnamefont {P.-G.}\ \bibnamefont {Reinhard}},\
  }\href {https://doi.org/10.1103/RevModPhys.75.121} {\bibfield  {journal}
  {\bibinfo  {journal} {Rev. Mod. Phys.}\ }\textbf {\bibinfo {volume} {75}},\
  \bibinfo {pages} {121} (\bibinfo {year} {2003})}\BibitemShut {NoStop}%
\bibitem [{\citenamefont {Vretenar}\ \emph {et~al.}(2005)\citenamefont
  {Vretenar}, \citenamefont {Afanasjev}, \citenamefont {Lalazissis},\ and\
  \citenamefont {Ring}}]{vretenar2005}%
  \BibitemOpen
  \bibfield  {author} {\bibinfo {author} {\bibfnamefont {D.}~\bibnamefont
  {Vretenar}}, \bibinfo {author} {\bibfnamefont {A.}~\bibnamefont {Afanasjev}},
  \bibinfo {author} {\bibfnamefont {G.}~\bibnamefont {Lalazissis}},\ and\
  \bibinfo {author} {\bibfnamefont {P.}~\bibnamefont {Ring}},\ }\href
  {https://doi.org/10.1016/j.physrep.2004.10.001} {\bibfield  {journal}
  {\bibinfo  {journal} {Phys. Rep.}\ }\textbf {\bibinfo {volume} {409}},\
  \bibinfo {pages} {101 } (\bibinfo {year} {2005})}\BibitemShut {NoStop}%
\bibitem [{\citenamefont {Robledo}\ \emph {et~al.}(2019)\citenamefont
  {Robledo}, \citenamefont {Rodr\'iguez},\ and\ \citenamefont
  {Rodr\'iguez-Guzm\'an}}]{robledo2019}%
  \BibitemOpen
  \bibfield  {author} {\bibinfo {author} {\bibfnamefont {L.~M.}\ \bibnamefont
  {Robledo}}, \bibinfo {author} {\bibfnamefont {T.~R.}\ \bibnamefont
  {Rodr\'iguez}},\ and\ \bibinfo {author} {\bibfnamefont {R.~R.}\ \bibnamefont
  {Rodr\'iguez-Guzm\'an}},\ }\href
  {http://stacks.iop.org/0954-3899/46/i=1/a=013001} {\bibfield  {journal}
  {\bibinfo  {journal} {J. Phys. G: Nucl. Part. Phys.}\ }\textbf {\bibinfo
  {volume} {46}},\ \bibinfo {pages} {013001} (\bibinfo {year}
  {2019})}\BibitemShut {NoStop}%
\bibitem [{\citenamefont {Schunck}(2019)}]{schunk2019}%
  \BibitemOpen
  \bibinfo {editor} {\bibfnamefont {N.}~\bibnamefont {Schunck}},\ ed.,\ \href
  {https://doi.org/10.1088/2053-2563/aae0ed} {\emph {\bibinfo {title} {Energy
  Density Functional Methods for Atomic Nuclei}}},\ 2053-2563\ (\bibinfo
  {publisher} {IOP Publishing},\ \bibinfo {year} {2019})\BibitemShut {NoStop}%
\bibitem [{\citenamefont {Rickey}\ \emph {et~al.}(1977)\citenamefont {Rickey},
  \citenamefont {Grau}, \citenamefont {Samuelson},\ and\ \citenamefont
  {Simms}}]{rickey1977}%
  \BibitemOpen
  \bibfield  {author} {\bibinfo {author} {\bibfnamefont {F.~A.}\ \bibnamefont
  {Rickey}}, \bibinfo {author} {\bibfnamefont {J.~A.}\ \bibnamefont {Grau}},
  \bibinfo {author} {\bibfnamefont {L.~E.}\ \bibnamefont {Samuelson}},\ and\
  \bibinfo {author} {\bibfnamefont {P.~C.}\ \bibnamefont {Simms}},\ }\href
  {https://doi.org/10.1103/PhysRevC.15.1530} {\bibfield  {journal} {\bibinfo
  {journal} {Phys. Rev. C}\ }\textbf {\bibinfo {volume} {15}},\ \bibinfo
  {pages} {1530} (\bibinfo {year} {1977})}\BibitemShut {NoStop}%
\bibitem [{\citenamefont {Leviatan}(1988)}]{leviatan1988}%
  \BibitemOpen
  \bibfield  {author} {\bibinfo {author} {\bibfnamefont {A.}~\bibnamefont
  {Leviatan}},\ }\href@noop {} {\bibfield  {journal} {\bibinfo  {journal}
  {Phys. Lett. B}\ }\textbf {\bibinfo {volume} {209}},\ \bibinfo {pages} {415}
  (\bibinfo {year} {1988})}\BibitemShut {NoStop}%
\bibitem [{\citenamefont {Alonso}\ \emph {et~al.}(1992)\citenamefont {Alonso},
  \citenamefont {Arias}, \citenamefont {Iachello},\ and\ \citenamefont
  {Vitturi}}]{alonso1992}%
  \BibitemOpen
  \bibfield  {author} {\bibinfo {author} {\bibfnamefont {C.}~\bibnamefont
  {Alonso}}, \bibinfo {author} {\bibfnamefont {J.}~\bibnamefont {Arias}},
  \bibinfo {author} {\bibfnamefont {F.}~\bibnamefont {Iachello}},\ and\
  \bibinfo {author} {\bibfnamefont {A.}~\bibnamefont {Vitturi}},\ }\href
  {https://doi.org/https://doi.org/10.1016/0375-9474(92)90235-C} {\bibfield
  {journal} {\bibinfo  {journal} {Nucl. Phys. A}\ }\textbf {\bibinfo {volume}
  {539}},\ \bibinfo {pages} {59} (\bibinfo {year} {1992})}\BibitemShut
  {NoStop}%
\bibitem [{\citenamefont {Iachello}\ \emph {et~al.}(2011)\citenamefont
  {Iachello}, \citenamefont {Leviatan},\ and\ \citenamefont
  {Petrellis}}]{iachello2011}%
  \BibitemOpen
  \bibfield  {author} {\bibinfo {author} {\bibfnamefont {F.}~\bibnamefont
  {Iachello}}, \bibinfo {author} {\bibfnamefont {A.}~\bibnamefont {Leviatan}},\
  and\ \bibinfo {author} {\bibfnamefont {D.}~\bibnamefont {Petrellis}},\
  }\href@noop {} {\bibfield  {journal} {\bibinfo  {journal} {Phys. Lett. B}\
  }\textbf {\bibinfo {volume} {705}},\ \bibinfo {pages} {379 } (\bibinfo {year}
  {2011})}\BibitemShut {NoStop}%
\bibitem [{\citenamefont {B\"oy\"ukata}\ \emph {et~al.}(2021)\citenamefont
  {B\"oy\"ukata}, \citenamefont {Alonso}, \citenamefont {Arias}, \citenamefont
  {Fortunato},\ and\ \citenamefont {Vitturi}}]{boyukata2021}%
  \BibitemOpen
  \bibfield  {author} {\bibinfo {author} {\bibfnamefont {M.}~\bibnamefont
  {B\"oy\"ukata}}, \bibinfo {author} {\bibfnamefont {C.~E.}\ \bibnamefont
  {Alonso}}, \bibinfo {author} {\bibfnamefont {J.~M.}\ \bibnamefont {Arias}},
  \bibinfo {author} {\bibfnamefont {L.}~\bibnamefont {Fortunato}},\ and\
  \bibinfo {author} {\bibfnamefont {A.}~\bibnamefont {Vitturi}},\ }\bibfield
  {journal} {\bibinfo  {journal} {Symmetry}\ }\textbf {\bibinfo {volume}
  {13}},\ \href {https://doi.org/10.3390/sym13020215} {10.3390/sym13020215}
  (\bibinfo {year} {2021})\BibitemShut {NoStop}%
\bibitem [{\citenamefont {Goriely}\ \emph {et~al.}(2009)\citenamefont
  {Goriely}, \citenamefont {Hilaire}, \citenamefont {Girod},\ and\
  \citenamefont {P\'eru}}]{D1M}%
  \BibitemOpen
  \bibfield  {author} {\bibinfo {author} {\bibfnamefont {S.}~\bibnamefont
  {Goriely}}, \bibinfo {author} {\bibfnamefont {S.}~\bibnamefont {Hilaire}},
  \bibinfo {author} {\bibfnamefont {M.}~\bibnamefont {Girod}},\ and\ \bibinfo
  {author} {\bibfnamefont {S.}~\bibnamefont {P\'eru}},\ }\href
  {https://doi.org/10.1103/PhysRevLett.102.242501} {\bibfield  {journal}
  {\bibinfo  {journal} {Phys. Rev. Lett.}\ }\textbf {\bibinfo {volume} {102}},\
  \bibinfo {pages} {242501} (\bibinfo {year} {2009})}\BibitemShut {NoStop}%
\bibitem [{\citenamefont {Nik\ifmmode \check{s}\else
  \v{s}\fi{}i\ifmmode~\acute{c}\else \'{c}\fi{}}\ \emph
  {et~al.}(2008)\citenamefont {Nik\ifmmode \check{s}\else
  \v{s}\fi{}i\ifmmode~\acute{c}\else \'{c}\fi{}}, \citenamefont {Vretenar},\
  and\ \citenamefont {Ring}}]{DDPC1}%
  \BibitemOpen
  \bibfield  {author} {\bibinfo {author} {\bibfnamefont {T.}~\bibnamefont
  {Nik\ifmmode \check{s}\else \v{s}\fi{}i\ifmmode~\acute{c}\else \'{c}\fi{}}},
  \bibinfo {author} {\bibfnamefont {D.}~\bibnamefont {Vretenar}},\ and\
  \bibinfo {author} {\bibfnamefont {P.}~\bibnamefont {Ring}},\ }\href
  {https://doi.org/10.1103/PhysRevC.78.034318} {\bibfield  {journal} {\bibinfo
  {journal} {Phys. Rev. C}\ }\textbf {\bibinfo {volume} {78}},\ \bibinfo
  {pages} {034318} (\bibinfo {year} {2008})}\BibitemShut {NoStop}%
\bibitem [{\citenamefont {Nomura}\ \emph {et~al.}(2017)\citenamefont {Nomura},
  \citenamefont {Rodr\'{\i}guez-Guzm\'an},\ and\ \citenamefont
  {Robledo}}]{nomura2017odd-3}%
  \BibitemOpen
  \bibfield  {author} {\bibinfo {author} {\bibfnamefont {K.}~\bibnamefont
  {Nomura}}, \bibinfo {author} {\bibfnamefont {R.}~\bibnamefont
  {Rodr\'{\i}guez-Guzm\'an}},\ and\ \bibinfo {author} {\bibfnamefont {L.~M.}\
  \bibnamefont {Robledo}},\ }\href {https://doi.org/10.1103/PhysRevC.96.064316}
  {\bibfield  {journal} {\bibinfo  {journal} {Phys. Rev. C}\ }\textbf {\bibinfo
  {volume} {96}},\ \bibinfo {pages} {064316} (\bibinfo {year}
  {2017})}\BibitemShut {NoStop}%
\bibitem [{\citenamefont {Bohr}(1953)}]{bohr1953}%
  \BibitemOpen
  \bibfield  {author} {\bibinfo {author} {\bibfnamefont {A.}~\bibnamefont
  {Bohr}},\ }\href@noop {} {\bibfield  {journal} {\bibinfo  {journal} {Mat.
  Fys. Medd. Dan. Vid. Selsk.}\ }\textbf {\bibinfo {volume} {27}},\ \bibinfo
  {pages} {16} (\bibinfo {year} {1953})}\BibitemShut {NoStop}%
\bibitem [{\citenamefont {Iachello}\ and\ \citenamefont {Arima}(1987)}]{IBM}%
  \BibitemOpen
  \bibfield  {author} {\bibinfo {author} {\bibfnamefont {F.}~\bibnamefont
  {Iachello}}\ and\ \bibinfo {author} {\bibfnamefont {A.}~\bibnamefont
  {Arima}},\ }\href@noop {} {\emph {\bibinfo {title} {The interacting boson
  model}}}\ (\bibinfo  {publisher} {Cambridge University Press, Cambridge},\
  \bibinfo {year} {1987})\BibitemShut {NoStop}%
\bibitem [{\citenamefont {Ring}\ and\ \citenamefont {Schuck}(1980)}]{RS}%
  \BibitemOpen
  \bibfield  {author} {\bibinfo {author} {\bibfnamefont {P.}~\bibnamefont
  {Ring}}\ and\ \bibinfo {author} {\bibfnamefont {P.}~\bibnamefont {Schuck}},\
  }\href@noop {} {\emph {\bibinfo {title} {The nuclear many-body problem}}}\
  (\bibinfo  {publisher} {Berlin: Springer-Verlag},\ \bibinfo {year}
  {1980})\BibitemShut {NoStop}%
\bibitem [{\citenamefont {Bohr}\ and\ \citenamefont
  {Mottelson}(1975{\natexlab{b}})}]{BM}%
  \BibitemOpen
  \bibfield  {author} {\bibinfo {author} {\bibfnamefont {A.}~\bibnamefont
  {Bohr}}\ and\ \bibinfo {author} {\bibfnamefont {B.~R.}\ \bibnamefont
  {Mottelson}},\ }\href@noop {} {\bibinfo {title} {Nuclear structure}}
  (\bibinfo {year} {1975}{\natexlab{b}})\BibitemShut {NoStop}%
\bibitem [{\citenamefont {{J. Decharge and M. Girod and D.
  Gogny}}(1975)}]{Gogny}%
  \BibitemOpen
  \bibfield  {author} {\bibinfo {author} {\bibnamefont {{J. Decharge and M.
  Girod and D. Gogny}}},\ }\href {https://doi.org/10.1016/0370-2693(75)90359-7}
  {\bibfield  {journal} {\bibinfo  {journal} {Phys. Lett. B}\ }\textbf
  {\bibinfo {volume} {55}},\ \bibinfo {pages} {361} (\bibinfo {year}
  {1975})}\BibitemShut {NoStop}%
\bibitem [{\citenamefont {Tian}\ \emph {et~al.}(2009)\citenamefont {Tian},
  \citenamefont {Ma},\ and\ \citenamefont {Ring}}]{tian2009}%
  \BibitemOpen
  \bibfield  {author} {\bibinfo {author} {\bibfnamefont {Y.}~\bibnamefont
  {Tian}}, \bibinfo {author} {\bibfnamefont {Z.~Y.}\ \bibnamefont {Ma}},\ and\
  \bibinfo {author} {\bibfnamefont {P.}~\bibnamefont {Ring}},\ }\href
  {https://doi.org/10.1016/j.physletb.2009.04.067} {\bibfield  {journal}
  {\bibinfo  {journal} {Phys. Lett. B}\ }\textbf {\bibinfo {volume} {676}},\
  \bibinfo {pages} {44 } (\bibinfo {year} {2009})}\BibitemShut {NoStop}%
\bibitem [{\citenamefont {Wilets}\ and\ \citenamefont {Jean}(1956)}]{gsoft}%
  \BibitemOpen
  \bibfield  {author} {\bibinfo {author} {\bibfnamefont {L.}~\bibnamefont
  {Wilets}}\ and\ \bibinfo {author} {\bibfnamefont {M.}~\bibnamefont {Jean}},\
  }\href {https://doi.org/10.1103/PhysRev.102.788} {\bibfield  {journal}
  {\bibinfo  {journal} {Phys. Rev.}\ }\textbf {\bibinfo {volume} {102}},\
  \bibinfo {pages} {788} (\bibinfo {year} {1956})}\BibitemShut {NoStop}%
\bibitem [{\citenamefont {Otsuka}\ \emph {et~al.}(1978)\citenamefont {Otsuka},
  \citenamefont {Arima},\ and\ \citenamefont {Iachello}}]{OAI}%
  \BibitemOpen
  \bibfield  {author} {\bibinfo {author} {\bibfnamefont {T.}~\bibnamefont
  {Otsuka}}, \bibinfo {author} {\bibfnamefont {A.}~\bibnamefont {Arima}},\ and\
  \bibinfo {author} {\bibfnamefont {F.}~\bibnamefont {Iachello}},\ }\href
  {https://doi.org/10.1016/0375-9474(78)90532-8} {\bibfield  {journal}
  {\bibinfo  {journal} {Nucl. Phys. A}\ }\textbf {\bibinfo {volume} {309}},\
  \bibinfo {pages} {1} (\bibinfo {year} {1978})}\BibitemShut {NoStop}%
\bibitem [{\citenamefont {Ginocchio}\ and\ \citenamefont
  {Kirson}(1980)}]{ginocchio1980}%
  \BibitemOpen
  \bibfield  {author} {\bibinfo {author} {\bibfnamefont {J.~N.}\ \bibnamefont
  {Ginocchio}}\ and\ \bibinfo {author} {\bibfnamefont {M.~W.}\ \bibnamefont
  {Kirson}},\ }\href {https://doi.org/10.1016/0375-9474(80)90387-5} {\bibfield
  {journal} {\bibinfo  {journal} {Nucl. Phys. A}\ }\textbf {\bibinfo {volume}
  {350}},\ \bibinfo {pages} {31} (\bibinfo {year} {1980})}\BibitemShut
  {NoStop}%
\bibitem [{\citenamefont {Caprio}\ and\ \citenamefont
  {Iachello}(2004)}]{caprio2004}%
  \BibitemOpen
  \bibfield  {author} {\bibinfo {author} {\bibfnamefont {M.~A.}\ \bibnamefont
  {Caprio}}\ and\ \bibinfo {author} {\bibfnamefont {F.}~\bibnamefont
  {Iachello}},\ }\href {https://doi.org/10.1103/PhysRevLett.93.242502}
  {\bibfield  {journal} {\bibinfo  {journal} {Phys. Rev. Lett.}\ }\textbf
  {\bibinfo {volume} {93}},\ \bibinfo {pages} {242502} (\bibinfo {year}
  {2004})}\BibitemShut {NoStop}%
\bibitem [{\citenamefont {Caprio}\ and\ \citenamefont
  {Iachello}(2005)}]{caprio2005}%
  \BibitemOpen
  \bibfield  {author} {\bibinfo {author} {\bibfnamefont {M.}~\bibnamefont
  {Caprio}}\ and\ \bibinfo {author} {\bibfnamefont {F.}~\bibnamefont
  {Iachello}},\ }\href
  {https://doi.org/https://doi.org/10.1016/j.aop.2005.02.003} {\bibfield
  {journal} {\bibinfo  {journal} {Annals of Physics}\ }\textbf {\bibinfo
  {volume} {318}},\ \bibinfo {pages} {454} (\bibinfo {year}
  {2005})}\BibitemShut {NoStop}%
\bibitem [{\citenamefont {Nomura}\ \emph {et~al.}(2008)\citenamefont {Nomura},
  \citenamefont {Shimizu},\ and\ \citenamefont {Otsuka}}]{nomura2008}%
  \BibitemOpen
  \bibfield  {author} {\bibinfo {author} {\bibfnamefont {K.}~\bibnamefont
  {Nomura}}, \bibinfo {author} {\bibfnamefont {N.}~\bibnamefont {Shimizu}},\
  and\ \bibinfo {author} {\bibfnamefont {T.}~\bibnamefont {Otsuka}},\ }\href
  {https://doi.org/10.1103/PhysRevLett.101.142501} {\bibfield  {journal}
  {\bibinfo  {journal} {Phys. Rev. Lett.}\ }\textbf {\bibinfo {volume} {101}},\
  \bibinfo {pages} {142501} (\bibinfo {year} {2008})}\BibitemShut {NoStop}%
\bibitem [{\citenamefont {Nomura}\ \emph {et~al.}(2012)\citenamefont {Nomura},
  \citenamefont {Shimizu}, \citenamefont {Vretenar}, \citenamefont {Nik\ifmmode
  \check{s}\else \v{s}\fi{}i\ifmmode~\acute{c}\else \'{c}\fi{}},\ and\
  \citenamefont {Otsuka}}]{nomura2012tri}%
  \BibitemOpen
  \bibfield  {author} {\bibinfo {author} {\bibfnamefont {K.}~\bibnamefont
  {Nomura}}, \bibinfo {author} {\bibfnamefont {N.}~\bibnamefont {Shimizu}},
  \bibinfo {author} {\bibfnamefont {D.}~\bibnamefont {Vretenar}}, \bibinfo
  {author} {\bibfnamefont {T.}~\bibnamefont {Nik\ifmmode \check{s}\else
  \v{s}\fi{}i\ifmmode~\acute{c}\else \'{c}\fi{}}},\ and\ \bibinfo {author}
  {\bibfnamefont {T.}~\bibnamefont {Otsuka}},\ }\href
  {https://doi.org/10.1103/PhysRevLett.108.132501} {\bibfield  {journal}
  {\bibinfo  {journal} {Phys. Rev. Lett.}\ }\textbf {\bibinfo {volume} {108}},\
  \bibinfo {pages} {132501} (\bibinfo {year} {2012})}\BibitemShut {NoStop}%
\bibitem [{\citenamefont {Scholten}(1985)}]{scholten1985}%
  \BibitemOpen
  \bibfield  {author} {\bibinfo {author} {\bibfnamefont {O.}~\bibnamefont
  {Scholten}},\ }\href
  {https://doi.org/https://doi.org/10.1016/0146-6410(85)90054-7} {\bibfield
  {journal} {\bibinfo  {journal} {Prog. Part. Nucl. Phys.}\ }\textbf {\bibinfo
  {volume} {14}},\ \bibinfo {pages} {189} (\bibinfo {year} {1985})}\BibitemShut
  {NoStop}%
\bibitem [{\citenamefont {Iachello}(1981)}]{IBFM-Book}%
  \BibitemOpen
  \bibinfo {editor} {\bibfnamefont {F.}~\bibnamefont {Iachello}},\ ed.,\
  \href@noop {} {\emph {\bibinfo {title} {Interacting Bose-Fermi Systems in
  Nuclei}}}\ (\bibinfo  {publisher} {Springer},\ \bibinfo {address} {New
  York},\ \bibinfo {year} {1981})\BibitemShut {NoStop}%
\bibitem [{\citenamefont {Nomura}\ \emph {et~al.}(2016)\citenamefont {Nomura},
  \citenamefont {Nik\ifmmode \check{s}\else \v{s}\fi{}i\ifmmode~\acute{c}\else
  \'{c}\fi{}},\ and\ \citenamefont {Vretenar}}]{nomura2016odd}%
  \BibitemOpen
  \bibfield  {author} {\bibinfo {author} {\bibfnamefont {K.}~\bibnamefont
  {Nomura}}, \bibinfo {author} {\bibfnamefont {T.}~\bibnamefont {Nik\ifmmode
  \check{s}\else \v{s}\fi{}i\ifmmode~\acute{c}\else \'{c}\fi{}}},\ and\
  \bibinfo {author} {\bibfnamefont {D.}~\bibnamefont {Vretenar}},\ }\href
  {https://doi.org/10.1103/PhysRevC.93.054305} {\bibfield  {journal} {\bibinfo
  {journal} {Phys. Rev. C}\ }\textbf {\bibinfo {volume} {93}},\ \bibinfo
  {pages} {054305} (\bibinfo {year} {2016})}\BibitemShut {NoStop}%
\bibitem [{\citenamefont {Otsuka}\ and\ \citenamefont {Yoshida}(1985)}]{PBOS}%
  \BibitemOpen
  \bibfield  {author} {\bibinfo {author} {\bibfnamefont {T.}~\bibnamefont
  {Otsuka}}\ and\ \bibinfo {author} {\bibfnamefont {N.}~\bibnamefont
  {Yoshida}},\ }\href@noop {} {} (\bibinfo {year} {1985}),\ \bibinfo {note}
  {{}JAERI-M (Japan Atomic Energy Research Institute) Report No.
  85}\BibitemShut {NoStop}%
\bibitem [{\citenamefont {Krane}\ and\ \citenamefont
  {Steffen}(1970)}]{krane1970}%
  \BibitemOpen
  \bibfield  {author} {\bibinfo {author} {\bibfnamefont {K.~S.}\ \bibnamefont
  {Krane}}\ and\ \bibinfo {author} {\bibfnamefont {R.~M.}\ \bibnamefont
  {Steffen}},\ }\href {https://doi.org/10.1103/PhysRevC.2.724} {\bibfield
  {journal} {\bibinfo  {journal} {Phys. Rev. C}\ }\textbf {\bibinfo {volume}
  {2}},\ \bibinfo {pages} {724} (\bibinfo {year} {1970})}\BibitemShut {NoStop}%
\bibitem [{\citenamefont {Lange}\ \emph {et~al.}(1982)\citenamefont {Lange},
  \citenamefont {Kumar},\ and\ \citenamefont {Hamilton}}]{lange1982}%
  \BibitemOpen
  \bibfield  {author} {\bibinfo {author} {\bibfnamefont {J.}~\bibnamefont
  {Lange}}, \bibinfo {author} {\bibfnamefont {K.}~\bibnamefont {Kumar}},\ and\
  \bibinfo {author} {\bibfnamefont {J.~H.}\ \bibnamefont {Hamilton}},\ }\href
  {https://doi.org/10.1103/RevModPhys.54.119} {\bibfield  {journal} {\bibinfo
  {journal} {Rev. Mod. Phys.}\ }\textbf {\bibinfo {volume} {54}},\ \bibinfo
  {pages} {119} (\bibinfo {year} {1982})}\BibitemShut {NoStop}%
\bibitem [{\citenamefont {Sambataro}\ \emph {et~al.}(1984)\citenamefont
  {Sambataro}, \citenamefont {Scholten}, \citenamefont {Dieperink},\ and\
  \citenamefont {Piccitto}}]{sambataro1984}%
  \BibitemOpen
  \bibfield  {author} {\bibinfo {author} {\bibfnamefont {M.}~\bibnamefont
  {Sambataro}}, \bibinfo {author} {\bibfnamefont {O.}~\bibnamefont {Scholten}},
  \bibinfo {author} {\bibfnamefont {A.}~\bibnamefont {Dieperink}},\ and\
  \bibinfo {author} {\bibfnamefont {G.}~\bibnamefont {Piccitto}},\ }\href
  {https://doi.org/https://doi.org/10.1016/0375-9474(84)90593-1} {\bibfield
  {journal} {\bibinfo  {journal} {Nucl. Phys. A}\ }\textbf {\bibinfo {volume}
  {423}},\ \bibinfo {pages} {333} (\bibinfo {year} {1984})}\BibitemShut
  {NoStop}%
\bibitem [{\citenamefont {Yoshida}\ and\ \citenamefont
  {Iachello}(2013)}]{yoshida2013}%
  \BibitemOpen
  \bibfield  {author} {\bibinfo {author} {\bibfnamefont {N.}~\bibnamefont
  {Yoshida}}\ and\ \bibinfo {author} {\bibfnamefont {F.}~\bibnamefont
  {Iachello}},\ }\href {https://doi.org/10.1093/ptep/ptt007} {\bibfield
  {journal} {\bibinfo  {journal} {Prog. Theor. Exp. Phys.}\ }\textbf {\bibinfo
  {volume} {2013}},\ \bibinfo {pages} {043D01} (\bibinfo {year}
  {2013})}\BibitemShut {NoStop}%
\bibitem [{\citenamefont {Hua}\ \emph {et~al.}()\citenamefont {Hua},
  \citenamefont {Guo},\ and\ \citenamefont {Petrache}}]{hua2020}%
  \BibitemOpen
  \bibfield  {author} {\bibinfo {author} {\bibfnamefont {W.}~\bibnamefont
  {Hua}}, \bibinfo {author} {\bibfnamefont {S.}~\bibnamefont {Guo}},\ and\
  \bibinfo {author} {\bibfnamefont {C.}~\bibnamefont {Petrache}},\ }\href@noop
  {} {\bibinfo {title} {to be published}},\ \bibinfo {howpublished}
  {{https://arxiv.org/abs/2011.14369}}\BibitemShut {NoStop}%
\bibitem [{\citenamefont {Urban}\ \emph {et~al.}(1984)\citenamefont {Urban},
  \citenamefont {Morek}, \citenamefont {Droste}, \citenamefont {Kotli\'nski},
  \citenamefont {Srebrny}, \citenamefont {Wrzesi\'nski},\ and\ \citenamefont
  {Stycze\'n}}]{urban1984}%
  \BibitemOpen
  \bibfield  {author} {\bibinfo {author} {\bibfnamefont {W.}~\bibnamefont
  {Urban}}, \bibinfo {author} {\bibfnamefont {T.}~\bibnamefont {Morek}},
  \bibinfo {author} {\bibfnamefont {C.}~\bibnamefont {Droste}}, \bibinfo
  {author} {\bibfnamefont {B.}~\bibnamefont {Kotli\'nski}}, \bibinfo {author}
  {\bibfnamefont {J.}~\bibnamefont {Srebrny}}, \bibinfo {author} {\bibfnamefont
  {J.}~\bibnamefont {Wrzesi\'nski}},\ and\ \bibinfo {author} {\bibfnamefont
  {J.}~\bibnamefont {Stycze\'n}},\ }\href {https://doi.org/10.1007/BF01881283}
  {\bibfield  {journal} {\bibinfo  {journal} {Z. Phys. A}\ }\textbf {\bibinfo
  {volume} {320}},\ \bibinfo {pages} {327} (\bibinfo {year}
  {1984})}\BibitemShut {NoStop}%
\bibitem [{\citenamefont {Bijker}\ and\ \citenamefont
  {Dieperink}(1982)}]{bijker1982}%
  \BibitemOpen
  \bibfield  {author} {\bibinfo {author} {\bibfnamefont {R.}~\bibnamefont
  {Bijker}}\ and\ \bibinfo {author} {\bibfnamefont {A.}~\bibnamefont
  {Dieperink}},\ }\href
  {https://doi.org/https://doi.org/10.1016/0375-9474(82)90391-8} {\bibfield
  {journal} {\bibinfo  {journal} {Nucl. Phys. A}\ }\textbf {\bibinfo {volume}
  {379}},\ \bibinfo {pages} {221} (\bibinfo {year} {1982})}\BibitemShut
  {NoStop}%
\bibitem [{\citenamefont {Arias}\ \emph {et~al.}(1986)\citenamefont {Arias},
  \citenamefont {Alonso},\ and\ \citenamefont {Lozano}}]{arias1986}%
  \BibitemOpen
  \bibfield  {author} {\bibinfo {author} {\bibfnamefont {J.~M.}\ \bibnamefont
  {Arias}}, \bibinfo {author} {\bibfnamefont {C.~E.}\ \bibnamefont {Alonso}},\
  and\ \bibinfo {author} {\bibfnamefont {M.}~\bibnamefont {Lozano}},\ }\href
  {https://doi.org/10.1103/PhysRevC.33.1482} {\bibfield  {journal} {\bibinfo
  {journal} {Phys. Rev. C}\ }\textbf {\bibinfo {volume} {33}},\ \bibinfo
  {pages} {1482} (\bibinfo {year} {1986})}\BibitemShut {NoStop}%
\bibitem [{\citenamefont {Heyde}\ and\ \citenamefont {Wood}(2011)}]{heyde2011}%
  \BibitemOpen
  \bibfield  {author} {\bibinfo {author} {\bibfnamefont {K.}~\bibnamefont
  {Heyde}}\ and\ \bibinfo {author} {\bibfnamefont {J.~L.}\ \bibnamefont
  {Wood}},\ }\href {https://doi.org/10.1103/RevModPhys.83.1467} {\bibfield
  {journal} {\bibinfo  {journal} {Rev. Mod. Phys.}\ }\textbf {\bibinfo {volume}
  {83}},\ \bibinfo {pages} {1467} (\bibinfo {year} {2011})}\BibitemShut
  {NoStop}%
\bibitem [{\citenamefont {Yao}\ \emph {et~al.}(2013)\citenamefont {Yao},
  \citenamefont {Bender},\ and\ \citenamefont {Heenen}}]{yao2013}%
  \BibitemOpen
  \bibfield  {author} {\bibinfo {author} {\bibfnamefont {J.~M.}\ \bibnamefont
  {Yao}}, \bibinfo {author} {\bibfnamefont {M.}~\bibnamefont {Bender}},\ and\
  \bibinfo {author} {\bibfnamefont {P.-H.}\ \bibnamefont {Heenen}},\ }\href
  {https://doi.org/10.1103/PhysRevC.87.034322} {\bibfield  {journal} {\bibinfo
  {journal} {Phys. Rev. C}\ }\textbf {\bibinfo {volume} {87}},\ \bibinfo
  {pages} {034322} (\bibinfo {year} {2013})}\BibitemShut {NoStop}%
\bibitem [{\citenamefont {Nomura}\ \emph {et~al.}(2013)\citenamefont {Nomura},
  \citenamefont {Rodr\'{\i}guez-Guzm\'an},\ and\ \citenamefont
  {Robledo}}]{nomura2013hg}%
  \BibitemOpen
  \bibfield  {author} {\bibinfo {author} {\bibfnamefont {K.}~\bibnamefont
  {Nomura}}, \bibinfo {author} {\bibfnamefont {R.}~\bibnamefont
  {Rodr\'{\i}guez-Guzm\'an}},\ and\ \bibinfo {author} {\bibfnamefont {L.~M.}\
  \bibnamefont {Robledo}},\ }\href {https://doi.org/10.1103/PhysRevC.87.064313}
  {\bibfield  {journal} {\bibinfo  {journal} {Phys. Rev. C}\ }\textbf {\bibinfo
  {volume} {87}},\ \bibinfo {pages} {064313} (\bibinfo {year}
  {2013})}\BibitemShut {NoStop}%
\end{thebibliography}%
\end{document}